\documentclass[aps, prl, twocolumn, superscriptaddress]{revtex4-2}
\usepackage{amsmath}
\usepackage{amssymb, times}
\usepackage{graphicx}
\usepackage{physics}
\usepackage{slashed}
\usepackage{mathtools}
\usepackage[colorlinks=true,linkcolor=blue,citecolor=blue,urlcolor=blue]{hyperref}

\newcommand{\beq}{\begin{equation}}
\newcommand{\eeq}{\end{equation}}

\DeclareMathOperator{\arsinh}{arsinh}

\newcommand{\NO}[1]{:\!#1\!:}

\def \tr {\operatorname{Tr}}

\def \A {\boldsymbol{A}}

\def \k {\mathbf{k}}
\def \q {\mathbf{q}} 
\def \p {\mathbf{p}}
\def \r {\mathbf{r}} 
\def \n {\mathbf{n}} 

\def \alphavec    {\boldsymbol{\alpha}}
\def \gammavec {\boldsymbol{\gamma}} 
\def \sigmavec   {\boldsymbol{\sigma}}

\def \ei {\hat{\mathbf e}_i}
\def \ej {\hat{\mathbf e}_j}

\def \H {\hat{H}}

\def\Hfg{\hat H_{f\mbox{\scriptsize-}g}}

\def\c{\hat{c}^{\vphantom{\dagger}}}
\def\cd{\hat{c}^\dagger}

\begin{document}
\title{Nonperturbative Bounds on the Ground-State Energy of Lattice QED}

\author{Pawe{\l} Zin}
\affiliation{National Centre for Nuclear Research, ul. Pasteura 7, PL-02-093 Warsaw, Poland}
\email{pawel.zin@ncbj.gov.pl}
\author{Andrzej Veitia}
\affiliation{Strzelecka 20, PL-58-560, Jelenia G\'ora, Poland}
\begin{abstract}
We establish nonperturbative bounds on the ground-state energy of lattice quantum electrodynamics with naive fermions in Coulomb gauge, within the Gaussian photon approximation. Our central result is a spectral comparison theorem showing that, in the thermodynamic limit, the minimum fermionic energy density over all link configurations coincides with that of the trivial gauge configuration. The proof exploits a logarithmic-determinant representation and dimensional reduction, without relying on reflection positivity, and applies to lattices of either parity. Combining this result with the positivity of the electromagnetic and Coulomb Hamiltonians, we derive an explicit lower bound on the interacting ground-state energy. In the limit \(\mu=ma\to0\), the ground-state energy per lattice site, relative to the Fock-vacuum energy, lies between approximately \(-0.2659\,q^2/a\) and zero. Finite-lattice counterexamples show that the thermodynamic limit is essential.
\end{abstract}

\maketitle
Quantum electrodynamics (QED) is among the most accurately tested theories in physics. Its perturbative formulation has produced predictions in remarkable agreement with experiment, including high-order calculations of the electron anomalous magnetic moment, which is now known through fifth order in the fine-structure constant $\alpha$  \cite{Volkov2019,Kitano2025}.

Despite these successes, fundamental questions concerning the mathematical foundations of perturbative QED remain unresolved. One of the most important is the convergence of the perturbative expansion. In his seminal 1952 paper, Dyson argued that the perturbation series in QED cannot converge \cite{Dyson1952}. Nevertheless, the increasing agreement between successive perturbative calculations and experimental measurements suggests that the series has an asymptotic character.

To the best of our knowledge, the asymptotic character of the QED perturbation series has not been rigorously established. For a divergent asymptotic series, successive perturbative approximations typically approach the exact result up to an optimal truncation order, beyond which adding further terms increases the error. Establishing such behavior requires understanding the relation between perturbative predictions and the full nonperturbative theory.

More fundamentally, the existence of a formal perturbative expansion does not guarantee that the underlying Hamiltonian is bounded below. Consider, for example, $\hat H_\epsilon=\hat p^2+\hat x^2+\epsilon \hat x^4$, with $\epsilon<0$. Although perturbation theory around the harmonic oscillator produces finite coefficients at every order, the full Hamiltonian is unbounded from below~\cite{BenderWu1969,Simon1970}. This illustrates why the existence of a ground state cannot be established by perturbation theory alone.

These questions motivate the study of nonperturbative formulations of QED. Lattice QED provides a natural framework for such investigations, since it introduces an ultraviolet cutoff while allowing the electromagnetic and fermionic degrees of freedom to be treated nonperturbatively.  Beyond its role as a regularization scheme, a discrete formulation also leaves open the possibility that the continuum description of space may cease to be valid at sufficiently short distances. Among the nonperturbative properties of QED,  the ground-state energy is a fundamental quantity through which the relationship between perturbative and nonperturbative descriptions can be explored. In a hypothetical isolated universe governed exclusively by QED, the absolute ground-state energy would have no observable significance since adding a constant to the Hamiltonian would not affect the dynamics. When gravity, for example, is included, vacuum energy can contribute to spacetime curvature, giving it observable physical consequences.

In this work, we derive nonperturbative upper and lower bounds on the ground-state energy density of lattice QED with naive fermions, formulated in Coulomb gauge within the Gaussian photon approximation. We obtain the lower bound by exploiting the spectral structure of the lattice fermion hopping Hamiltonian, without invoking reflection positivity, thereby avoiding complications associated with boundary terms and lattice parity. The upper bound is established using the variational principle. Importantly, both bounds are established within the charge-neutral sector, as required by Gauss’s law on a finite lattice with periodic boundary conditions. Both bounds are extensive, scaling linearly with the number of lattice sites, and remain well defined in the thermodynamic limit. With the Fock-vacuum energy chosen as zero, both bounds vanish as $\alpha\to0$ at fixed lattice spacing, after taking the thermodynamical limit. This establishes the weak-coupling agreement of the ground-state and Fock-vacuum energy densities, providing a necessary consistency check for perturbation theory about the Fock vacuum. The detailed spectral proof, finite-lattice counterexamples, and evaluation of the lattice sums are presented in the Supplemental Material~\cite{SM}.

\section{The Lattice QED Hamiltonian}
\begin{figure}[t!]
 \centering
\includegraphics[width=0.4\textwidth]{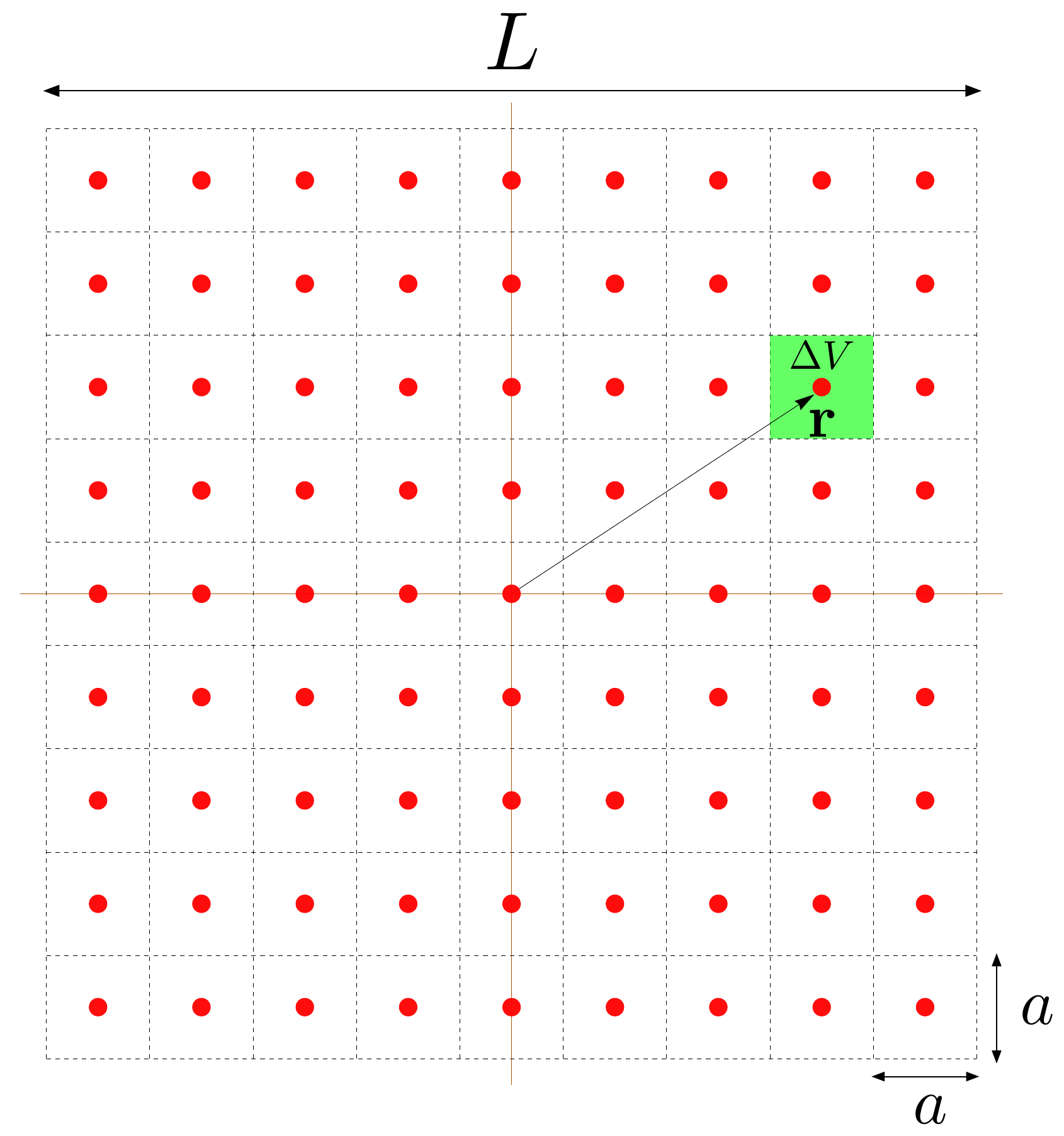}
\caption{Spatial Lattice $\Omega_N $. A  two-dimensional cross section of the three-dimensional regular cubic lattice is shown. The red points denote lattice sites. There are $2N+1$ sites in each spatial direction, with lattice spacing $a$ and linear extent $L=a(2N+1)$. The shaded square represents the cross section of an elementary cubic cell of volume $\Delta V=a^3$. Hence, the total physical volume is $ V=(2N+1)^3\Delta V=L^3$.
We denote lattice sites by $\r\in\Omega_N.$}
\label{fig:lattice_1}
\end{figure}

We consider a cubic spatial lattice with periodic boundary
conditions, lattice spacing $a$, and sites
\beq
\Omega_N =\left\{
\mathbf r=a\, \mathbf j \mid 
\mathbf j\in\{-N,\ldots,N\}^3
\right\}.
\eeq
The lattice contains $N_\Omega=(2N+1)^3$ sites, with
linear extent $L=a(2N+1)$ and total volume
$V=N_\Omega\Delta V=L^3$, where $\Delta V=a^3$, as described in Fig.~\ref{fig:lattice_1}. The corresponding set of allowed momenta is
\begin{equation}
\Omega_N^\star=\big\{
\k=\frac{2\pi}{L}\mathbf n \mid 
\mathbf n\in\{-N,\ldots,N\}^3
\big\}.
\end{equation}
Having defined the spatial lattice and its allowed momenta, we now introduce the mode expansions of the Dirac and transverse electromagnetic fields in Coulomb gauge. At $t=0$, these take the form
\begin{eqnarray}
\label{eq:dirac_field_expansion}
\hat{\Psi}_{\alpha}(0, \r)&=& \frac{1}{\sqrt{V}}\sum_{\k, s}\sqrt{\frac{m}{E_k}}\big(u_{\alpha}(\bar \k, s)  e^{i \k \cdot \r} \hat{b}^{}_{\k, s} \nonumber \\
&+& v_{\alpha}(\bar \k, s)e^{-i \k \cdot \r} \hat{d}^{\dagger}_{\k, s}  \big), 
\end{eqnarray} 
and 
\beq
\label{eq:A_field_expansion}
\hat{A}_i (0, \r) = \sideset{}{'}\sum_{\mathbf k, \sigma}\frac{1}{\sqrt{2\omega_\k V}}\big(\epsilon_i(\k, \sigma)e^{i \k \cdot (\r + \frac{a}{2} \ei)} \hat{a}_{\k, \sigma} + \mathrm{h.c.}\big).
\eeq
Here, $\bar{k}_i=a^{-1}\sin(ak_i)$ and $\tilde{k}_i=2a^{-1}\sin(ak_i/2)$ denote the lattice momenta associated with the fermionic and electromagnetic fields, respectively. The corresponding energies are $E_{\k}=\sqrt{m^2+\bar{\k}^{2}}$ and $\omega_{\k}=|\tilde{\k}|$. In both field expansions, the momentum sums extend over \(\Omega_N^\star\), with the prime indicating the exclusion of the zero-momentum photon mode. The photon creation and annihilation operators satisfy the canonical bosonic commutation relations, while the corresponding fermionic operators obey the canonical anticommutation relations,
\begin{eqnarray}
[\hat a^{}_{\k, \sigma}, \hat a^{\dagger}_{\p, \sigma'}] &=& \delta_{\k, \p} \delta_{\sigma, \sigma'}\\
\{\hat b^{}_{\k, s}, \hat b^{\dagger}_{\p, s'}\} &=&  \{\hat d^{}_{\k, s}, \hat d^{\dagger}_{\p, s'}\} = \delta_{\k, \p} \delta_{s, s'}. 
\end{eqnarray} 
All other commutators and anticommutators within the respective bosonic and fermionic sectors vanish. Operators belonging to different sectors commute. The bi-spinors $u({\k},s)$ and $v({\k},s)$, $s=1, 2,$ correspond to the positive- and negative-energy plane-wave solutions of the free lattice Dirac equation, respectively. Their explicit expressions and normalization conventions are given in the Supplemental Material~\cite{SM}. The vectors $\boldsymbol\epsilon(\mathbf k,\sigma)$ , with $\sigma=1,2$, describe the transverse photon polarizations and satisfy the transversality condition
\beq
\tilde{\k}\cdot\boldsymbol\epsilon(\k,\sigma)=0, 
\label{eq:transversality_cond_1}
\eeq 
and the completeness relation, 
\beq
\sum_{\sigma=1}^{2}
\epsilon^{\phantom{}}_i(\k,\sigma)
\epsilon_j^*(\k,\sigma) = \delta_{ij} - \frac{\tilde{k}_{i} \tilde{k}_j}{\tilde{k}^2}. 
\label{eq:completness_pol}
\eeq
The transversality condition (\ref{eq:transversality_cond_1}) ensures that the electromagnetic field satisfies the lattice Coulomb-gauge constraint,
\beq
\nabla^-\cdot\hat{\mathbf A}(0,\r)=0,
\eeq
where $\nabla_i^-f(\r)=[f(\r)-f(\ r-a\ei)]/a)$ denotes the backward lattice derivative, and $\ei$ is the unit vector along the $i$-th spatial direction. Similarly, the forward lattice derivative is defined by $\nabla_i^+f(\r)=[f(\r+a\ei)-f(\r)]/a$.

We consider a Hamiltonian formulation of lattice QED with naive fermions, closely related to the Kogut–Susskind construction~\cite{KogutSusskind1975}. Working in Coulomb gauge, we eliminate the longitudinal electromagnetic degrees of freedom and retain the Gaussian approximation for the transverse photon Hamiltonian. Related Coulomb-gauge lattice formulations have been investigated in Refs.~\cite{Li2025Coulomb}. In the Gaussian photon approximation, the lattice QED Hamiltonian consists of four contributions: the free transverse electromagnetic Hamiltonian, the fermionic mass term, the gauge-covariant hopping Hamiltonian, and the instantaneous Coulomb interaction. These are given, respectively, by
\begin{align}
\label{eq:Hem}
\hat H_{\mathrm{em}}^{(0)}
&=\sideset{}{'}\sum_{\mathbf k,\sigma}
\omega_{\mathbf k}
\big(
\hat a_{\k,\sigma}^\dagger
\hat a^{}_{\k,\sigma}+\frac{1}{2}
\big),\\
\hat H_m
&=m\Delta V\sum_{\mathbf r}
\hat\Psi^\dagger(\mathbf r)\beta\hat\Psi(\mathbf r),\\
\Hfg 
&=-i\frac{\Delta V}{2a}\sum_{\mathbf r,i}
\Big[
\hat\Psi^\dagger(\mathbf r)\alpha^i
\hat\Psi(\mathbf r+a\hat{\mathbf e}_i)
\hat U_i^\dagger(\mathbf r)\notag\\
&\hspace{2.5cm}
-\hat\Psi^\dagger(\mathbf r+a\hat{\mathbf e}_i)
\alpha^i\hat\Psi(\mathbf r)
\hat U_i(\mathbf r)
\Big],\\
\hat H_c &= -\frac{q^2}{2} (\Delta V)^2  \sum_{\r, \r'} \hat J_0 (\r)\, G(\r - \r')\, \hat{J}_0 (\r'). 
\label{eq:Hc_0}
\end{align}
Unless otherwise indicated, all spatial lattice sums extend over $\r\in\Omega_N$, while momentum sums extend over $\k\in\Omega_N^\star$. Here, $\hat J_0(\r) = \hat \Psi^{\dagger} (\r) \hat \Psi^{}(\r)$  denotes the fermionic density operator, while $\hat U_i(\r)$ is the unitary link variable associated with the directed link from $\r$ to $\r+a\ei$. The link operator is explicitly given by
\beq
\hat U_i (\r) = e^{- i q a \hat A_i (\r)}, 
\eeq
where $q= - |e|$ denotes the electron charge. The function $G(\r-\r')$  is the Green function of the lattice Laplacian $\Delta=\sum_i\nabla_i^-\nabla_i^+ = \sum_i\nabla_i^+\nabla_i^-$, with the spatial zero mode removed. The Green function is given by 
\beq
G(\r-\r') =  -\frac{1}{V} \sideset{}{'}\sum_{\k \in \Omega_N^{\star}}\frac{e^{i \k \cdot (\r - \r')}}{\tilde{\k}^2}. 
\label{eq:def_Green}
\eeq
Since the zero-momentum mode is excluded from Eq.~(\ref{eq:def_Green}), the Green function satisfies
\beq
\sum_{\r \in \Omega_N} G(\r - \r') = \sum_{\r' \in \Omega_N}G(\r - \r')= 0. 
\label{eq:Green_1}
\eeq
Since the Fock vacuum is translationally invariant, the expectation value of the charge density is independent of 
$\r$. Consequently, Eq. (\ref{eq:Green_1}) allows us to replace the charge-density operators in the Coulomb Hamiltonian by their normal-ordered counterparts without changing the operator. Thus,
\beq
\hat H_c =-\frac{q^2}{2} (\Delta V)^2  \sum_{\r, \r'} \NO{\hat J_0 (\r)} G(\r - \r') \NO{\hat{J}_0 (\r')}. 
\label{eq:Hc_1}
\eeq
This expression is the lattice counterpart of the Coulomb interaction written in terms of normal-ordered charge densities in Bjorken and Drell~\cite{BjorkenDrell1965}. We emphasize that Eq. (\ref{eq:Hc_1}) does not amount to normal ordering the Coulomb Hamiltonian itself. In general, normal ordering an interacting Hamiltonian can generate additional lower-order operator terms and thereby modify its spectrum, rather than merely shifting the energy levels by a constant \cite{BenderWu1971}. This distinction is particularly important here, since our prescription preserves the positivity of the Coulomb Hamiltonian. 

We introduce the lattice Fourier transform of the
fermion-density operator,
\begin{equation}
\hat J_0(\k)
:=\Delta V\sum_{\r}
\hat J_0(\r)e^{-i\k\cdot\r}.
\end{equation}
Substituting Eq. (\ref{eq:def_Green}) into Eq.~(\ref{eq:Hc_1}), we obtain
\begin{equation}
\hat H_c=\frac{q^2}{2V}
\sideset{}{'}\sum_{\k}
\frac{\NO{\hat J_0^\dagger(\k)}
      \NO{\hat J_0(\mathbf k)}}
{\tilde{\k}^{\,2}}
\geq 0.
\label{eq:Hc_2}
\end{equation}
The physical electric charge density is $q:\hat J_0(\mathbf r):$, where normal ordering is defined with respect to the free fermionic vacuum. Accordingly, the operator lattice Poisson equation reads \(\Delta\hat A_0(\mathbf r)=-q \NO{\hat J_0(\mathbf r)}\). Since the lattice Laplacian annihilates constant functions, periodic boundary conditions require the global charge-neutrality constraint
\beq
\hat Q \ket{\mathrm{phys}} = 0, \qquad \hat Q = \Delta V\sum_{\r} \NO{\hat J_0(\r)}.
\label{eq:charge_neutrality_cond}
\eeq
Finally, we choose the Fock vacuum $\ket {0} =\ket{0_f}\otimes \ket{0_g}$  as our reference state, where $\ket{0_f}$  and $\ket{0_g}$ denote the fermionic and gauge-field Fock vacua, respectively. Defining \(\hat H_{\mathrm{QED}}\) as the sum of the four contributions in Eqs. (\ref{eq:Hem})--(\ref{eq:Hc_0}), we subtract its Fock-vacuum expectation value,
\beq
\hat H_{\mathrm{QED}} \rightarrow \hat H_{\mathrm{QED}} - \bra{0}\hat H_{\mathrm{QED}}\ket{0}.
\label{eq:Fock_sub}
\eeq
This subtraction changes only the energy reference and leaves the dynamics unchanged. Consequently, our shifted Hamiltonian satisfies 
\beq
\bra{0} \hat H_{\mathrm{QED}} \ket{0}=0. 
\eeq
\section{A simple Upper bound on the ground-state energy}
Let $E_\Omega$ denote the ground-state energy of the
interacting lattice QED Hamiltonian in the charge-neutral
sector. Since $\langle0|\hat H_{\mathrm{QED}}|0\rangle=0$,
the variational principle immediately gives
\beq
E_\Omega\leq 0.
\eeq
To obtain a strict inequality, consider the normalized
trial state
\beq
\ket{\Phi_\epsilon}=
\frac{\ket{0}+\epsilon \ket{\chi}}
{\sqrt{1+|\epsilon|^2+
2\Re[\epsilon\langle0|\chi\rangle]}},
\eeq
where $\epsilon\in\mathbb C$ and $\ket{\chi}$ is a
normalized state in the charge-neutral sector.
Its energy expectation value is
\beq
E_{\Phi_\epsilon}=
\frac{
2\operatorname{Re}[
\epsilon\langle0|\hat H_{\mathrm{QED}}|\chi\rangle]
+|\epsilon|^2
\langle\chi|\hat H_{\mathrm{QED}}|\chi\rangle
}{
1+|\epsilon|^2+
2\operatorname{Re}[\epsilon\langle0|\chi\rangle]
}.
\eeq
For a state satisfying
$\bra{0}\hat H_{\mathrm{QED}}\ket{\chi}\neq 0$,
we can choose the phase of $\epsilon$ such that
the linear term in the numerator is negative.
For sufficiently small $|\epsilon|$, this term
dominates the quadratic contribution. Therefore,
\beq
E_\Omega<0.
\eeq
Such a state is relatively easy to construct. For example, one may choose $\ket{\chi}=\ket{0_f}\otimes
\ket{0_{g,s}}$, where \(|0_{g,s}\rangle\) denotes a suitably squeezed photon vacuum. For nonzero coupling, the gauge-covariant hopping term can provide the required nonvanishing matrix element.
\section{Lower bound on the ground-state energy}

We begin by combining the fermionic mass and
gauge-covariant hopping terms into a single
fermionic Hamiltonian,
\beq
\hat H_f := \hat H_m+\Hfg. 
\label{eq:fermionic_ham}
\eeq
Since the Coulomb Hamiltonian is positive semidefinite, as shown in Eq.~(\ref{eq:Hc_2}), and the free electromagnetic Hamiltonian is bounded from below by its Fock-vacuum energy, we have
\begin{eqnarray}
\hat H_c &\geq& 0\\
\hat H^{(0)}_{\mathrm{em}} - \bra{0_g}\hat H^{(0)}_{\mathrm{em}}\ket{0_g}
&=& \sideset{}{'}\sum_{\k, \sigma}\omega_{\k} \hat a^{\dagger}_{\k, \sigma}\hat a^{} _{\k, \sigma}\geq 0. 
\end{eqnarray}
Combining these equations with the Fock-vacuum energy subtraction~(\ref{eq:Fock_sub}), we obtain the operator inequality
\beq
\hat H_{\mathrm{QED}} \geq \hat{H}_f - E_F I,  
\eeq
where $E_{F} = \bra{0}\hat H_c\ket{0} + \bra{0} \Hfg \ket{0} + \bra{0}\hat H_m \ket{0}.$
Consequently, the ground-state energy of the interacting theory satisfies
\beq
E_{\Omega} \geq \min_U E_f^{(0)}[U] - E_F, 
\label{eq:lower_bound_intro}
\eeq
where $E_f^{(0)}[U]$ denotes the ground-state energy of the fermionic Hamiltonian for a fixed configuration of  link variables $U = \{U_i (\r)\}$, restricted to the charge-neutral sector.  For a fixed configuration of link variables $U$,
the fermionic Hamiltonian (\ref{eq:fermionic_ham}) can be expressed in terms
of a single-particle Hermitian matrix $h[U]$ as
\begin{equation}
\hat H_f[U]=\Delta V\sum_{\mathbf r,\mathbf r'}
\hat\Psi^\dagger(\mathbf r)
h_{\mathbf r,\mathbf r'}[U]
\hat\Psi(\mathbf r'),
\end{equation}
where
\begin{align}
h_{\mathbf r,\mathbf r'}[U]
={}&m\delta_{\mathbf r,\mathbf r'}\beta
-\frac{i}{2a}\sum_{j=1}^3\alpha^j
\Big[
U_j^*(\mathbf r)
\delta_{\mathbf r',\mathbf r+a\hat{\mathbf e}_j}
\nonumber\\
&\hspace{1.2cm}
-U_j(\mathbf r-a\hat{\mathbf e}_j)
\delta_{\mathbf r',\mathbf r-a\hat{\mathbf e}_j}
\Big].
\end{align}
The matrix \(h[U]\) acts on the \(4N_\Omega\)-dimensional single-particle Hilbert space. Its spectral properties determine the fermionic ground-state energy \(E_f^{(0)}[U]\).  As shown in the Supplemental Material~\cite{SM},
the spectrum of $h[U]$ is symmetric about zero.
Together with the charge-neutrality constraint,
this implies
\begin{equation}
E_f^{(0)}[U]=-\frac{1}{2}\Tr|h[U]|, 
\label{eq:h_norm}
\end{equation}
where $|h[U]| = \sqrt{h[U]^2}.$

Our main result follows from a spectral comparison
based on a logarithmic-determinant representation
of the fermionic ground-state energy. By applying
a dimensional-reduction inequality, we eliminate
the dependence on the gauge-field configuration
up to a correction that vanishes per lattice site
in the thermodynamic limit. As shown in the
Supplemental Material~\cite{SM}, we obtain
\begin{equation}
\lim_{N_\Omega\to\infty}
\frac{
\min_U E_f^{(0)}[U]-E_f^{(0)}[I]
}{N_\Omega}=0.
\label{eq:main_result}
\end{equation}
Here, $I$ denotes the trivial link configuration,
$U_i(\mathbf r)=1$. Importantly, this result holds
independently of lattice parity, and does not
require the trivial configuration to minimize
the fermionic energy at finite volume. For the trivial link configuration $U=I$, the single-particle Hamiltonian has eigenvalues \(\pm E_{\mathbf k}\), each doubly degenerate. Consequently, Eq. (\ref{eq:h_norm}) gives
\beq
E_f^{(0)}[I] = -\frac{1}{2} \Tr\abs{h[I]} = -2 \sum_{\k \in \Omega_N^{\star}} E_{\k}. 
\label{eq:E_config_I}
\eeq
Combining this with Eqs.~(\ref{eq:lower_bound_intro}) and (\ref{eq:main_result}), we obtain the following lower bound in the thermodynamic limit:
\beq
\lim_{N_{\Omega}\to \infty} \frac{E_{\Omega}}{N_{\Omega}} \geq 
\lim_{N_{\Omega}\to \infty} \frac{E_f^{(0)}[I] - E_F}{N_{\Omega}}. 
\label{eq:main_result_1}
\eeq
Using the Fock-vacuum expectation values calculated in the Supplemental Material~\cite{SM},  the mass and hopping contributions combine into
\beq
\bra{0} \hat H_f \ket{0} = -2 \sum_{\k \in \Omega_N^{\star}}\frac{m^2 + \mathcal U(N) \;\bar \k^2}{E_{\k}}, 
\eeq
where $\mathcal{U}(N) = \bra{0_{g}}\hat U_i (\r) \ket{0_g}$. Combining this expression with Eq.~(\ref{eq:E_config_I}) and the definition of \(E_F\), we obtain
\beq
E^{(0)} _f[I] - E_F =-\bra{0}\hat H_c\ket{0} -2 \Big(1 - \mathcal{U}(N)\Big)\times \sum_{\k \in \Omega_N^{\star}}\frac{\bar \k^2}{E_{\k}}. 
\eeq
The remaining quantities can be expressed in terms of
four dimensionless lattice sums, evaluated in the
Supplemental Material~\cite{SM}. Specifically, 
\begin{align}
\langle0|\hat H_c|0\rangle
&=\frac{q^2N_\Omega}{8a}
\big[C_0(N)-C_1(N,\mu)\big],\\
\mathcal U (N)
&=\exp\big[-\frac{q^2}{12}C_2(N)\big],\\
C_3(N,\mu)
&=\frac{2a}{N_\Omega}
\sum_{\mathbf k\in\Omega_N^\star}
\frac{\bar{\mathbf k}^{\,2}}{E_{\mathbf k}},
\label{eq:C3_main}
\end{align}
where $\mu=ma$. 
In the thermodynamic limit, the lattice sums evaluated
in the Supplemental Material~\cite{SM} yield
\begin{align}
C_0(\infty)&\approx 1.01092,
& C_1(\infty,0)&\approx 0.33313,\\
C_2(\infty)&\approx 0.9107,
& C_3(\infty,0)&\approx 2.3876.
\end{align}
Combining Eqs.~(\ref{eq:main_result})--(\ref{eq:C3_main}), and using
$1-e^{-x}\leq x$ for $x\geq 0$, we obtain,
in the limit $\mu\to 0$,
\begin{equation}
\lim_{N_\Omega\to\infty}
\frac{E_\Omega}{N_\Omega}
\geq -C\frac{q^2}{a},
\end{equation}
where
\begin{equation}
C=\frac{C_0(\infty)-C_1(\infty,0)}{8}
+\frac{C_2(\infty)C_3(\infty,0)}{12}
\approx 0.2659.
\end{equation}

In summary, we have established nonperturbative bounds on the ground-state energy of lattice QED in the charge-neutral sector, within the Gaussian photon approximation. The central ingredient is a spectral comparison theorem showing that the trivial gauge configuration is asymptotically optimal for the fermionic energy density, even though this need not hold at finite volume. Our approach avoids reflection positivity and applies to lattices of either parity. In particular, the cancellation of the free-fermion contribution ensures that the lower bound vanishes in the weak-coupling limit, consistently with our choice of the Fock vacuum as the reference state. A natural extension of our work would be to investigate whether similar bounds can be established for other lattice fermion discretizations, such as Wilson or staggered fermions \cite{rothe2012lattice}.
\clearpage

\bibliographystyle{apsrev4-2}
\bibliography{references}
\clearpage
\onecolumngrid

\setcounter{equation}{0}
\renewcommand{\theequation}{S\arabic{equation}}
\renewcommand{\theHequation}{supplement.\arabic{equation}}

\begin{center}
{\large\bfseries Supplemental Material}
\end{center}

\section{ Dirac spinor conventions} 
We adopt the conventions and notation of Ref.~\cite{ItzyksonZuber}, using the West Coast metric signature $g^{\mu\nu}=\operatorname{diag}(1,-1,-1,-1)$. In the Dirac representation, we have
\beq
\label{Dirac_rep_matrices}
\gamma^{0} = \beta,\quad \gammavec = \beta \alphavec, 
\eeq
where
\beq
\beta = \begin{bmatrix} 
I & 0 \\
0 & -I
\end{bmatrix},\quad \alphavec =  
\begin{bmatrix} 
0 & \sigmavec \\
\sigmavec & 0
\end{bmatrix}.
\eeq
Here,  $\sigmavec = (\sigma^1, \sigma^2, \sigma^3)$ denotes the standard $2\times 2$ Pauli matrices. 
These matrices obey $\{\sigma^{i}, \sigma^{j}\} = 2\delta^{i j} I.$ Consequently, the gamma matrices in Eq.~(\ref{Dirac_rep_matrices}) indeed satisfy the Clifford algebra property
\beq
\{\gamma^{\mu}, \gamma^{\nu} \} = 2 g^{\mu \nu} I. 
\eeq
In addition, we have 
\beq
{\gamma^{\mu}}^{\dagger} = \gamma^{0}\gamma^{\mu}\gamma^{0}. 
\eeq
The standard solutions of Dirac's equation in momentum space are
\begin{eqnarray} 
\label{eq:udef}
u(\k, s) &=& \frac{\slashed{k} + m}{\sqrt{2m (m + E_k)}} u^{(s)}(m, 0)\\
\label{eq:vdef}
v(\k, s) &=& \frac{-\slashed{k} + m}{\sqrt{2m (m + E_k)}} v^{(s)}(m, 0), 
\end{eqnarray}
where $E_{\mathbf k}=\sqrt{\mathbf k^2+m^2}$,  $k^\mu=(E_{\mathbf k},\mathbf k)$ and 
\beq
u^{(1)}(m, 0) = \begin{bmatrix}
1\\
0 \\ 
0 \\
0 
\end{bmatrix},\quad 
u^{(2)}(m, 0) = \begin{bmatrix}
0\\
1 \\ 
0 \\
0 
\end{bmatrix},\quad
v^{(1)}(m, 0) = \begin{bmatrix}
0\\
0 \\ 
1 \\
0 
\end{bmatrix},\quad
v^{(2)}(m, 0) = \begin{bmatrix}
0\\
0 \\ 
0\\
1 
\end{bmatrix}.
\eeq
With this normalization, the spinors satisfy 
\begin{eqnarray}
\label{eq:u_norm_1}
\bar u(\k,s)u(\k,s') &=& \delta_{ss'},\\
\label{eq:v_norm_2}
\bar v(\k,s)v(\k,s') &=& -\delta_{ss'}.
\end{eqnarray}
The spinors also satisfy the following identities, which will be used throughout this work:
\begin{eqnarray}
\label{ident_dirac_11}
\sum_{s_1, s_2} \abs{u^{\dagger}(\k_1, s_1) v(\k_2, s_2)}^2 &=& \frac{E_{\k_1} E_{\k_2} + \k_1 \cdot \k_2 - m^2}{m^2},\\
\sum_{s} v^{\dagger}(\k, s) \alpha^{i} v(\k, s) &=& \frac{2 k^i}{m}. 
\label{ident_dirac_2}
\end{eqnarray}

\section{Lower Bound on the Ground-State Energy}
\subsection{Properties of the Single-Particle Hamiltonian} 
We first combine the mass and hopping terms
\beq
\H_f = \H_m + \Hfg. 
\eeq
We then write this Hamiltonian as 
\beq
\H_f = \Delta V \sum_{\r, \r '} \hat \Psi^{\dagger}(\r)\, h_{\r, \r'}[\hat U]\,\hat \Psi(\r'),  
\label{eq:Hf_appendix}
\eeq
where 
\beq
h_{\r, \r'}[\hat U] = m  \delta_{\r, \r'} \beta- \frac{i}{2a} \sum_{i=1}^{3} 
\big[ \delta_{\r', \r+ a \ei} \hat U_i^{\dagger}(\r)\ - \delta_{\r' , \r - a \ei}\hat U_i (\r - a \ei)\big]\alpha^i. 
\label{eq:single_ham_appendix_1}
\eeq

Since
\beq
[\hat A_i (\r), \hat A_j(\r')]=0, 
\eeq
we can introduce their common eigenstates, 
\beq
\hat A_i (\r) \ket{\A} = A_i (\r) \ket{\A}, 
\eeq
and consequently, 
\beq
\hat U_i (\r)\ket{\A} = U_i (\r) \ket{\A}. 
\eeq
In the basis $\ket{\A}$  the operator-valued matrix $h[\hat U]$ reduces, for each fixed gauge-field configuration $\A$, to an ordinary $4N_\Omega\times4N_\Omega$ Hermitian matrix $h[U].$
To investigate the properties of $h[U]$, it is convenient to introduce the orthonormal single-particle basis
\beq
\ket{\r, s}, \qquad \r \in \Omega_N,\qquad  s=1, \ldots, 4, 
\eeq
where $\r$ labels the lattice sites and $s$ denotes the Dirac component.  
We introduce the gauge-covariant lattice translation operators
\beq
T_j [U] = \sum_{\r\in \Omega_N} U^{*}_j (\r) \ketbra{\r}{\r + a \ej}.
\eeq
In terms of these operators, the single-particle Hamiltonian (\ref{eq:single_ham_appendix_1}) can be written as
\beq
h[U] = m I_{\Omega_N} \otimes \beta + K[U], 
\label{eq:single_ham_appendix_2}
\eeq
where 
\beq
K[U] = - \frac{i}{2a} \sum_{j=1}^3 \big(T^{}_j [U] - T^{\dagger}_j [U]\big) \otimes \alpha^j, 
\label{eq:K[U]_def}
\eeq
is the  hopping matrix. Because of the periodic boundary conditions, the matrices $T_j [U]$ are unitary. Moreover, since the Dirac matrices satisfy $\{\alpha^j, \beta\}=0$, the hopping matrix obeys 
\beq
\{K[U], I_{\Omega_N}\otimes \beta\}=0. 
\label{eq:K_1}
\eeq
Consequently, 
\beq
h[U]^2 = m^2 I_{4 N_{\Omega}} + K[U]^2. 
\label{eq:h[U]_squared}
\eeq
In addition, since $\beta$ is invertible, Eq.~(\ref{eq:K_1}) implies that the eigenvalues of $K[U]$ occur in opposite pairs. Indeed, if 
\beq
K[U]\, \mathbf v = \kappa\, \mathbf v,  
\eeq
then 
\beq
K[U] \big(I_{\Omega_N}\otimes \beta\big)\, \mathbf v = - \kappa (I_{\Omega_N}\otimes \beta)\,\mathbf v. 
\eeq
Thus, the spectrum of $K[U]$ is symmetric about zero. We now show that the spectrum of the full single-particle Hamiltonian $h[U]$ is also symmetric about zero. To this end, we introduce the matrix
\beq
\Gamma := \alpha^1 \alpha^2 \alpha^3 \beta = i \begin{bmatrix} 
0 & -I\\
I & 0
\end{bmatrix}. 
\eeq
The Clifford algebra implies
\beq
\{\Gamma,\alpha^j\}=\{\Gamma,\beta\}=0,
\eeq
and therefore
\beq
\{\Gamma,h[U]\}=0.
\eeq
Since $\Gamma$ is invertible, the eigenvalues of $h[U]$ occur in opposite pairs with equal multiplicities. Together with Eq. (\ref{eq:h[U]_squared}), this spectral symmetry implies that, for $m>0$, the single-particle Hamiltonian has exactly $2N_\Omega$ positive and $2N_\Omega$ negative eigenvalues. Before proceeding to the fermionic many-body Hamiltonian, we collect some useful properties of the hopping matrix that will be needed to establish the lower bound. These properties follow directly from the structure of the hopping matrix and hold independently of the gauge-field configuration.

The unitarity of the covariant lattice translations also provides a uniform bound on the operator norm of $K[U]$: 
\beq
\norm{K[U]} = \sqrt{\lambda_{\max}\Big[K[U]^{\dagger} K[U]\Big]} \leq \frac{1}{2a} \sum_{j=1}^3 \norm{T_j [U] - T_j^{\dagger}[U]} \norm{\alpha^j}\leq \frac{3}{a}. 
\label{eq:Knorm}
\eeq
We next evaluate the trace of $h[U]^2$ using the algebraic properties established above. Using Eqs.~(\ref{eq:K[U]_def}) and (\ref{eq:h[U]_squared}),  together with $
\Tr(\alpha^i\alpha^j)=4\delta^{ij}$, we obtain
\beq
\Tr[h[U]^2]
= 4m^2N_\Omega
-\frac{1}{a^2}\sum_{j=1}^{3}
\Tr
\left[(T_j[U]-T_j^\dagger[U])^2\right].
\eeq
Since the covariant lattice translations are unitary, we have
\beq
(T_j[U]-T_j^\dagger[U])^2
=
T_j[U]^2+(T_j^\dagger[U])^2-2I_{\Omega_N}.
\eeq
Moreover, a translation by two lattice sites has no fixed points on our periodic lattice, implying
\beq
\Tr_j[U]^2
=\Tr[(T_j^\dagger[U])^2]=0.
\eeq
Therefore, 
\beq
\Tr[h[U]^2] = 4 N_{\Omega} \big( m^2 + \frac{3}{2 a^2}\big). 
\label{eq:trace_h[U]_squared}
\eeq
Note that this result is independent of the gauge configuration $U.$

\subsection{The Neutral Fermionic Ground-State Energy}
For each fixed gauge-field configuration $U$, the Hermitian matrix $h[U]$ admits an orthonormal basis of eigenvectors satisfying
\beq
h[U]v_j[U]=\lambda_j[U]v_j[U],
\qquad j=1,\ldots,4N_\Omega.
\label{eq:eigenvalue_1}
\eeq
Consider now 
\beq
v_j( \r;U)=\braket{\r}{v_j [U]} \in \mathbb{C}^4.  
\eeq
The orthonormality of the eigenvectors implies
\beq
\sum_r v_j^{\dagger}(\r; U) v^{}_k (\r; U) = \delta_{j k}. 
\label{eq:orthonormal_1}
\eeq
We now introduce the fermionic annihilation operators associated with the single-particle eigenvectors,
\beq
\c_j[U] := \sqrt{\Delta V}
\sum_{\r}
v_j^\dagger(\r;U)\hat\Psi(\r).
\label{eq:c_def}
\eeq
Using the CAR of the fermionic fields, together with Eq.~(\ref{eq:orthonormal_1}), we obtain 
\beq
\{\c_j [U], \cd_k[U] \} = \delta_{j k}, \qquad \{\c_j [U], \c_k [U]\}=0. 
\eeq
The inverse transformation is 
\beq
\hat \Psi (\r) = \frac{1}{\sqrt{\Delta V}}\sum_{j=1}^{4 N_{\Omega}} v_j (\r ; U) \c_j [U]. 
\eeq
Substituting this relation into the Hamiltonian (\ref{eq:Hf_appendix}), and using the eigenvalue equation (\ref{eq:eigenvalue_1}), we obtain 
\beq
\hat H_f [U] = \sum_{j=1}^{4 N_{\Omega}}\lambda_j [U]\, \cd_j[U] \c_j[U]. 
\label{eq:Hf_2}
\eeq

We now impose the charge-neutrality constraint. Recall that the total charge operator is 
\beq
\hat Q = \Delta V \sum_{\r \in \Omega_N } \NO{\hat \Psi^{\dagger} (\r) \hat \Psi(\r)} = \Delta V \sum_{\r \in \Omega_N } \hat \Psi^{\dagger}(\r) \hat \Psi (\r) - 2 N_{\Omega} I. 
\eeq
In terms of the operators (\ref{eq:c_def}), we have 
\beq
\hat Q = \sum_{j=1}^{4 N_{\Omega}} \cd_j [U] \c_j[U] - 2 N_{\Omega} I. 
\eeq
The charge-neutrality condition 
\beq
\hat Q \ket{\Phi} =0 
\eeq
therefore requires 
\beq
\sum_{j=1}^{4 N_{\Omega}} \cd_j [U] \c_j [U] \ket{\Phi}= 2 N_{\Omega} \ket{\Phi}. 
\eeq
We introduce the occupation-number operators
\beq
\hat n_j[U]:=\cd_j [U] \c_j[U].
\eeq
The canonical anticommutation relations imply
\beq
\hat n_j [U]^2 = \hat n_j [U], \qquad [\hat n_j [U], \hat n_k [U]] = 0,  
\eeq
for all $j$ and $k.$ Thus, all occupation-number operators admit a simultaneous eigenbasis, 
\beq
\hat n_j [U] \ket{\mathbf n, U} = n_j \ket{\mathbf n, U}, \qquad n_j \in \{0, 1\}. 
\eeq
In this basis, the fermionic Hamiltonian (\ref{eq:Hf_2}) is diagonal, with eigenvalues 
\beq
E_f [\mathbf n, U] = \sum_{j=1}^{4 N_{\Omega}} n_j \,\lambda_j [U]. 
\eeq
Moreover, the charge neutrality condition becomes 
\beq
\sum_{j=1}^{4 N_{\Omega}} n_j = 2 N_{\Omega}. 
\eeq
We order the eigenvalues of the single-particle Hamiltonian as
\beq
\lambda_1[U]\leq\lambda_2[U]\leq\cdots\leq\lambda_{4N_\Omega}[U].
\eeq
Since charge neutrality requires exactly $2N_\Omega$ occupied fermionic modes, the minimum energy is obtained by occupying the $2N_\Omega$ lowest single-particle levels. Therefore, the minimum fermionic energy in the charge-neutral sector is
\beq
E_f^{(0)}[U]=\sum_{j=1}^{2N_\Omega}\lambda_j[U]. 
\eeq
Using the spectral symmetry of $h[U]$  established in the previous subsection, the minimum fermionic energy can be expressed as 
\beq
E_f^{(0)}[U]=-\frac{1}{2} \sum_{j=1}^{4 N_{\Omega}} |\lambda_j [U]| = -\frac{1}{2} \tr[\sqrt{h[U]^{\dagger} h[U]}] = -  \frac{1}{2} \norm{h[U]}_1, 
\label{eq:norm_1} 
\eeq
where $\norm{\cdot}_1$ denotes the trace norm. 
\subsection{The Cauchy--Schwarz Bound}
We first derive a simple lower bound on the neutral fermionic ground-state energy using the Cauchy–Schwarz inequality: 
\beq
\Big( \sum_{j=1}^{4 N_{\Omega}} \abs{\lambda_j}\Big)^2 \leq \Big(\sum_{j=1}^{4 N_{\Omega}} 1^2 \Big) \Big(\sum_{j=1}^{4 N_{\Omega}} \abs{\lambda_j}^2 \Big). 
\eeq
Then, from Eq.~(\ref{eq:norm_1}), we immediately find that 
\beq
E_f^{(0)}[U] \geq - \frac{1}{2} \sqrt{ 4 N_{\Omega} \Tr[h[U]^2]}
\eeq
Substituting the trace identity (\ref{eq:trace_h[U]_squared}), we finally obtain 
\beq
E_f^{(0)}[U] \geq -2 N_{\Omega} \sqrt{m^2 + \frac{3}{2 a^2}}. 
\label{eq:cauchy_schwarz_bound}
\eeq
As we shall show below, this bound can be attained on a $4\times4\times4$ lattice for a suitable gauge-field configuration. However, it cannot be attained in the thermodynamic limit, 
$N_{\Omega} \rightarrow \infty.$ This motivates the search for a sharper lower bound that exploits additional properties of the single-particle Hamiltonian.
\subsection{The Log-Determinant Method}
We now develop a sharper lower bound by expressing the neutral fermionic ground-state energy in terms of a logarithmic determinant. To this end, for $s>0$, we introduce the matrix
\beq
D[s; U] := s I_{4 N_{\Omega}} + i K[U], \qquad s> 0. 
\label{eq:D_matrix_1}
\eeq
First, since $K[U]$ is Hermitian,
\beq
D[s; U] D[s; U]^{\dagger} = s^2 I_{4 N_{\Omega}} + K[U]^2. 
\eeq
Moreover, since the eigenvalues of $K[U]$ occur in opposite pairs, the determinant of $D[s;U]$ is strictly positive:
\beq
\det D[s;U] =\prod_{j=1}^{4N_\Omega}(s+i\kappa_j[U]) =\prod_{j=1}^{2N_\Omega}
\left(s^2+\kappa_j[U]^2\right)>0,
\eeq
where $\kappa_j [U]$, $j=1, \ldots 4 N_{\Omega}$ denote the eigenvalues of $K[U]$, ordered so that $k_{j +2 N_{\Omega}}[U] = - \kappa_j [U],$  $j=1, \ldots, 2 N_{\Omega}.$
From the above expressions, we obtain 
\beq
\log \det D[s; U] = \frac{1}{2} \Tr \log(s^2 I_{4 N_{\Omega}} + K[U]^2). 
\label{eq:log_det_0}
\eeq
To relate the logarithmic determinant to the fermionic ground-state energy, we use the integral identity
\beq
\int_0^\infty
\log\left(\frac{t^2+x}{t^2+y}\right)\,dt
=
\pi(\sqrt{x}-\sqrt{y}),
\qquad x,y>0.
\eeq
Applying this identity  to the eigenvalues of $K[U]^2$, with $x=m^2+\kappa_j[U]^2$ and $y=m^2,$ and summing over $j$, we obtain
\beq
\sum_{j=1}^{4N_\Omega}
\left(\sqrt{m^2+\kappa_j[U]^2}-m\right)
=
\frac{2}{\pi}\int_0^\infty
\log\frac{\det D[\sqrt{t^2+m^2};U]}
{(t^2+m^2)^{2N_\Omega}}\,dt.
\eeq
Combining this with Eqs.~(\ref{eq:norm_1}) and (\ref{eq:h[U]_squared}) gives
\beq
E_f^{(0)}[U] = - 2 m N_{\Omega} - \frac{1}{\pi}\int_{0}^{\infty} \log \frac{ 
\det D[\sqrt{t^2+m^2};  U]}{(t^2 +m^2)^{2 N_{\Omega}}} dt. 
\label{eq:integral_rep_1}
\eeq
Note that this identity holds for every link configuration $U$ on the finite lattice $\Omega_N$.
\subsection{Dimensional Reduction}
We now establish a dimensional-reduction inequality for the logarithmic determinant appearing in Eq.~(\ref{eq:integral_rep_1}). To make the recursive structure explicit, we consider a periodic lattice in arbitrary spatial dimension $d$, with the three-dimensional case recovered by setting $d=3$. We consider a $d$-dimensional periodic lattice with $M$ sites along each spatial direction, where $M\geq3$ is an arbitrary integer. The lattice sites are labeled by
\beq
\mathbf n=(n_1,\ldots,n_d),
\qquad n_j=1,\ldots,M,
\eeq
with all site indices understood modulo $M$. The total number of lattice sites is $M^d$. We retain the four-component Dirac spinor structure and introduce the Hilbert space
\beq
\mathcal H_d=\mathbb C^{M^d}\otimes\mathbb C^4,
\qquad d=1,2,3,
\eeq
equipped with the orthonormal basis
\beq
\ket{\n, \mu}, \qquad \mu=1,\ldots, 4. 
\eeq
For each spatial direction $j=1,\ldots,d$,  we introduce the gauge-covariant lattice translations
\beq
T_j^{(d)}[U] = \sum_{\mathbf n}
U_j^*(\mathbf n)
|\mathbf n\rangle
\langle\mathbf n+\mathbf e_j|,
\eeq
where $\ej$  is the unit vector along the $j$-th spatial direction. We then define the $d$-dimensional hopping matrix
\beq
K_d [U] = - \frac{i}{2 a}  \sum_{j=1}^d \big(T_j^{(d)}[U] - T_j^{(d)}[U]^{\dagger} \big)\otimes \alpha^j. 
\label{eq:K_d}
\eeq
Here $K_d [U]$ acts on $\mathcal{H}_d$ so its dimension is $4 M^d \times 4 M^d.$ In analogy with Eq.~(\ref{eq:D_matrix_1}), we consider the matrix 
\beq
D_d[s ; U]:= s I_{4 M^d} + i K_d [U], \qquad s>0. 
\label{eq:Dd_1}
\eeq
We now establish the reduction from $d$  to $d-1$ spatial dimensions. To this end, we partition the lattice into $M$ layers perpendicular to the $d$-th direction. By writing
\beq
\n=(\n_\perp, n),
\qquad
\n_\perp=(n_1,\ldots,n_{d-1}),
\label{eq:perp_decomp}
\eeq
each layer is a $d-1$-dimensional periodic lattice containing $M^{d-1}$ sites. An arbitrary vector in $\mathcal{H}_d$ can be written as 
\beq
\ket{\Phi} = \sum_{\n} \sum_{\mu} \Phi_{\mu}(\n) \ket{\n} \otimes \ket{\mu}. 
\eeq
Using the decomposition (\ref{eq:perp_decomp}), we can equivalently write
\beq
\ket{\Phi} = \sum_{n=1}^M \ket{\Phi_n}\otimes \ket{n}, 
\eeq
where the layer vectors are defined by
\beq
\ket{\Phi_n} = \sum_{\n_{\perp}} \Phi_{\mu}(\n_{\perp}, n) \ket{\n_\perp} \otimes \ket{\mu} \in \mathcal{H}_{d-1}.
\eeq
Each layer vector has therefore 
\beq
d_{\perp} = 4 M^{d-1}
\eeq
components. Acting with $D_d[ s ; U]$ on $\ket{\Phi}\in \mathcal H_d$, we can write the result in terms of its layer components 
\beq
D_d[s ; U] \ket{\Phi} = \sum_{n=1}^{M} \ket{(D_d \Phi)_n} \otimes \ket{n}. 
\eeq
Substituting Eq.~(\ref{eq:Dd_1}) and collecting the results associated with each layer, we obtain 
\beq
\ket{(D_d \Phi)_n} = A_n \ket{\Phi_n} + C_n \ket{\Phi_{n+1}} - C^\dagger_{n-1} \ket{\Phi_{n-1}}. 
\label{eq:layers_exp}
\eeq
Here, $A_n$ and $C_n$ are $d_\perp\times d_\perp$ matrices describing the hopping within and between neighboring layers, respectively. Explicitly, the diagonal blocks are given by 
\beq
A_n = s I_{d_{\perp}} + i K_{d-1}[U^{(n)}], 
\label{eq:An_appendix}
\eeq
where $U^{(n)}$ denotes the link configuration within the $n$-th layer. To describe the coupling between neighboring layers, we introduce the matrix
\beq
J=I_{M^{d-1}}\otimes\alpha^d,
\label{eq:J_appendix}
\eeq
and 
\beq
\mathcal U_n=
\sum_{\n_\perp}
U_d(\n_\perp,n)
\ket{\mathbf n_\perp}\bra{\mathbf n_\perp}
\otimes I_4.
\eeq
Then 
\beq
C_n =\frac{1}{2 a} \mathcal{U}_n^{\dagger} J. 
\eeq
Since the link variables have unit modulus and $J^2=I_{d_\perp}$, it follows that
\beq
C_n^{\dagger} C_n = \frac{1}{4 a^2} I_{d_{\perp}}. 
\eeq
Consequently, the interlayer coupling matrices are invertible, with
\beq
C_n^{-1} = 4 a^2 C_n^{\dagger}. 
\eeq
In the basis $\bigl\{
\ket{\n_{\perp}, \mu} \otimes\ket{n}
\bigr\}_{n,\n_\perp,\mu}$, ordered by the layer index, the equation (\ref{eq:layers_exp}) gives the following cyclic block-tridiagonal representation of $D_d[s;U]$:
\beq
D_d[s;U]=
\begin{bmatrix}
A_1 & C_1 & 0 & \cdots & -C_M^\dagger \\
-C_1^\dagger & A_2 & C_2 & \cdots & 0 \\
0 & -C_2^\dagger & A_3 & \ddots & \vdots \\
\vdots & \vdots & \ddots & \ddots & C_{M-1} \\
C_M & 0 & \cdots & -C_{M-1}^\dagger & A_M
\end{bmatrix}
\eeq
The determinant of this matrix can be expressed in terms of transfer matrices \cite{Molinari2008}: 
\beq 
\det D_d[s;U]
=
(2a)^{-4M^d}
\left|
\det\left(
I_{2d_\perp}
-\mathcal T_M\mathcal T_{M-1}\cdots\mathcal T_1
\right)
\right|,
\label{eq:determinant}
\eeq
where 
\beq
\mathcal T_n=
\begin{bmatrix}
-C_n^{-1}A_n & C_n^{-1}C_{n-1}^{\dagger}\\
I_{d_\perp} & 0
\end{bmatrix}, \qquad n=1, \ldots, M, 
\eeq
with the periodic convention $C_0=C_M$.
\subsection{The Singular-Value Bound}
We now derive an upper bound on the determinant in Eq.~(\ref{eq:determinant}) using the singular values of the transfer matrices.
For an arbitrary $n\times n$ complex matrix $A$, we define
\beq
v(A):=\prod_{j=1}^{n}\max\{1,\sigma_j(A)\},
\eeq
where $\sigma_j(A),$  $j=1,\ldots,n$, denote the singular values of $A$. The quantity $v(A)$ satisfies the inequalities \cite{HornJohnson1991}
\beq
|\det(I_n - A)| \leq 2^{\dim A} v(A), \qquad v(A B) \leq v(A) v(B). 
\eeq
Both hold for arbitrary $n\times n$ complex matrices $A$ and $B.$ Applying these inequalities to the transfer-matrix representation (\ref{eq:determinant}), we obtain
\beq
\det D_d [s; U] \leq (2 a) ^{-4 M_d} 2^{2 d_{\perp}} v(T_{M}\cdots T_1) \leq (2 a) ^{-4 M_d} 2^{2 d_{\perp}}\prod_{n=1}^{M} v(T_n). 
\eeq
Taking logarithms, we find 
\beq
\log \det D_{d}[s; U] \leq - 4 M^{d} \log(2 a) + 2 d_{\perp} \log 2 + \sum_{n=1}^M \log v(T_n). 
\label{eq:log_det_ineq_1}
\eeq
To evaluate $v(\mathcal T_n)$, we introduce the matrix
\beq
B_n=2a J\,A_n, 
\eeq
where $A_n$ and $J$ are defined in Eqs.~(\ref{eq:An_appendix}) and (\ref{eq:J_appendix}). Since the Clifford algebra implies
\beq
\{J,K_{d-1}[U^{(n)}]\}=0,
\eeq
and $J^2=I_{d_\perp}$, we obtain
\beq
B_n^\dagger=B_n,
\qquad
B_n^2=4a^2\big(s^2I_{d_\perp}+K_{d-1}[U^{(n)}]^2\big).
\label{eq:B_squared}
\eeq
These identities allow us to determine the singular values of the transfer matrices. To this end, we first express $\mathcal T_n$ as a product of matrices whose singular values are readily obtained: 
\beq
\mathcal T_n=
\begin{bmatrix}
\mathcal U_n&0\\
0&I_{d_\perp}
\end{bmatrix}
\begin{bmatrix}
-B_n&I_{d_\perp}\\
I_{d_\perp}&0
\end{bmatrix}
\begin{bmatrix}
I_{d_\perp}&0\\
0&\mathcal U_{n-1}
\end{bmatrix}.
\eeq
Since the outer matrices are unitary, the singular values of $\mathcal T_n$ coincide with those of the central matrix
\beq
\mathcal B = \begin{bmatrix}
-B_n&I_{d_\perp}\\
I_{d_\perp}&0
\end{bmatrix}
\eeq
Since $B_n$ is Hermitian, let $b_j, j=1,\ldots,d_\perp$, denote its eigenvalues. In the corresponding eigenbasis, the matrix $\mathcal B$ decomposes into $2\times2$ blocks of the form
\beq
\begin{bmatrix}
-b_j&1\\
1&0
\end{bmatrix}. 
\label{eq:block}
\eeq
Since $\mathcal B$ is also Hermitian, its singular values are the absolute values of its eigenvalues. Each $2\times2$ block in Eq.~(\ref{eq:block}) has determinant -1. Its two eigenvalues are therefore real, nonzero, and of opposite signs, with absolute values whose product is one. Thus, the singular values occur in reciprocal pairs $\sigma_j,\sigma_j^{-1}$, where $\sigma_j\geq1$ satisfy
\beq
\sigma_j-\sigma_j^{-1}=|b_j|,
\qquad
\log\sigma_j=\arsinh \frac{|b_j|}{2}.
\eeq
Consequently, 
\beq 
v(\mathcal T_{n})
=\prod_{j=1}^{d_\perp}
\max\{1,\sigma_j\}
\max\{1,\sigma_j^{-1}\}\\
=\prod_{j=1}^{d_\perp}\sigma_j,
\eeq
Taking logarithms and using Eq.~(\ref{eq:B_squared}), we obtain 
\beq
\log v(\mathcal{T}_n) = \Tr \arsinh\big(a \sqrt{s^2 I_{d_\perp}+ K_{d-1}[U^{n}]^2}\big). 
\eeq
Finally, substituting this expression into the inequality (\ref{eq:log_det_ineq_1}) gives 
\beq
\log \det D_d[s; U] \leq - 4 M^d \log (2 a) + 2 d_{\perp} \log 2 + \sum_{n=1}^M  \Tr \arsinh\big(a \sqrt{s^2 I_{d_\perp}+ K_{d-1}[U^{(n)}]^2}\big).
\label{eq:log_det_bound_dimensional}
\eeq
Importantly, this bound is independent of the longitudinal link variables $U_d(\mathbf n_\perp,n)$. It depends only on the gauge configurations $U^{(n)}$ within the $d-1$-dimensional layers.
\subsection{The Dimensional-Reduction Inequality}
We now use the singular-value bound (\ref{eq:log_det_bound_dimensional}) to establish a recursive inequality relating the logarithmic determinants in $d$ and $d-1$  spatial dimensions. To express the right-hand side of Eq.~(\ref{eq:log_det_bound_dimensional}) in terms of lower-dimensional logarithmic determinants, we use the integral identity
\beq
\arsinh(a\sqrt{x})-\log(2a)
=
\frac{1}{2}\int_{-\pi}^{\pi}
\log\left(x+\frac{\sin^2p}{a^2}\right)
\frac{dp}{2\pi},
\qquad x>0.
\eeq
Applying this identity to the trace in Eq.~(\ref{eq:log_det_bound_dimensional}) gives 
\beq
\log\det D_d[s;U]\leq 2d_\perp\log2+\frac12\sum_{n=1}^{M}\int_{-\pi}^{\pi}\operatorname{Tr}\log\left[\Big(s^2+\frac{\sin^2p}{a^2}\right)I_{d_\perp}+K_{d-1}[U^{(n)}]^2\Big]\frac{dp}{2\pi}.
\label{eq:almost_there}
\eeq
Since $K_{d-1}[U^{(n)}]$ anticommutes with $J$, its spectrum is symmetric about zero. We introduce the parameter
\beq
s_p := \sqrt{s^2 + \frac{\sin^2 p}{a^2}}. 
\eeq
Using the same spectral argument that led to Eq.~(\ref{eq:log_det_0}), we obtain
\beq
\log\det D_{d-1}[s_p;\, U^{(n)}] = \frac1 2 \Tr\log\Big(s_p^2I_{d_\perp}+K_{d-1}[U^{(n)}]^2\Big). 
\eeq
Substituting this relation into Eq.~(\ref{eq:almost_there}) we arrive at the sought recursive relation 
\beq
\log\det D_d[s;U]\leq
2d_\perp\log2+
\sum_{n=1}^{M}\int_{-\pi}^{\pi}
\log\det D_{d-1}[s_p;U^{(n)}]\frac{dp}{2\pi}.
\label{eq:dimensional_rec_appendix}
\eeq
This inequality allows us to systematically reduce the problem from three spatial dimensions to zero, through the successive steps 
\beq
3\to2\to1\to0, 
\eeq
 while keeping track of all contributions at each stage. To complete the recursion, we introduce the terminal case d=0, corresponding to a single lattice site with four spinor components and no hopping terms. Thus,
\beq
K_0=0,\qquad D_0[s]=sI_4,
\label{eq:terminal_case}
\eeq
and consequently,
\beq
\log\det D_0[s]=4\log s.
\eeq
At each reduction step, the correction in Eq.~(\ref{eq:dimensional_rec_appendix}) is $2d_\perp\log2$. Accounting for the multiplicity of the lower-dimensional layers, we have
\beq
2M^k d_\perp^{(d-k)}\log2=8M^{d-1}\log2,
\eeq
independently of $k$. Since the recursion involves $d$ steps, the total correction is $8dM^{d-1}\log2$. Specializing to $d=3$ and iterating Eq.~(\ref{eq:dimensional_rec_appendix}) down to the terminal case (\ref{eq:terminal_case}), we obtain
\beq
\log\det D_3[s;U]\leq +24M^2\log2 + 2M^3\int_{[-\pi,\pi]^3}\log\Big(s^2+\frac{1}{a^2}\sum_{j=1}^{3}\sin^2p_j\Big)\frac{d^3\p}{(2\pi)^3}
\label{eq:finally_appendix}
\eeq
This bound holds for every gauge-field configuration and for both even and odd lattice sizes. The entire gauge dependence has been eliminated, leaving a universal momentum integral and a finite-size correction that grows only with the surface area of the lattice. More importantly, the momentum integral in Eq.~(\ref{eq:finally_appendix}) is directly related to the spectrum of the free single-particle $h[I]$. This connection allows us to express the bound in terms of the free fermionic ground-state energy.
Indeed, for the free-field configuration $U=I$, the eigenvalues of the hopping matrix $K[I]$ are
\beq
\kappa_{\n}^{\pm}
=
\pm\frac{1}{a}
\sqrt{\sum_{j=1}^{3}\sin^2\Big(\frac{2\pi n_j}{M}\Big)},
\eeq
where $n_j=1,\ldots,M$, and each eigenvalue is doubly degenerate. Consequently, 
\beq
\log\det D_3[s;I]
=
2\sum_{n_1,n_2,n_3=1}^{M}
\log\Big(s^2+\frac{1}{a^2}\sum_{j=1}^{3}\sin^2\frac{2\pi n_j}{M}\Big).
\eeq
Dividing by $M^3$ and taking the thermodynamic limit, the sum becomes a Riemann integral:
\beq
\lim_{M\to\infty}\frac{\log\det D_3[s;I]}{M^3}
=
2\int_{[-\pi,\pi]^3}
\log\bigg(s^2+\frac{1}{a^2}\sum_{j=1}^{3}\sin^2p_j\bigg)
\frac{d^3\p}{(2\pi)^3}.
\eeq
Combining this with Eq.~(\ref{eq:finally_appendix}), and noting that the finite-size correction vanishes after division by $M^3$, we obtain
\beq
\lim_{M\to\infty}\sup_{U} 
\frac{\log\det D_3[s;U]}{M^3}
\leq
\lim_{M\to\infty}
\frac{\log\det D_3[s;I]}{M^3}.
\eeq
Since $U=I$ is one of the configurations included in the supremum,
\beq
\sup_U\log\det D_3[s;U]
\geq \log\det D_3[s;I].
\eeq
Thus, 
\beq
\lim_{M\to\infty}\max_U
\frac{\log\det D_3[s;U]}{M^3}
=
\lim_{M\to\infty}
\frac{\log\det D_3[s;I]}{M^3}.
\label{eq:thermodynamical_1}
\eeq
Finally, introduce the quantity 
\beq
Q_M(s)=\max_U\frac{1}{M^3}
\log\frac{\det D[s;U]}{\det D[s;I]}.
\label{eq:ratio}
\eeq
Equation (\ref{eq:thermodynamical_1}) implies 
\beq
\lim_{M\to\infty}Q_M(s)=0,\qquad s>0.
\label{eq:Qto0}
\eeq
Here and in what follows, we drop the subscript 3 from $D_3[s;U]$, since we have returned to the physical three-dimensional lattice. Moreover, since $U=I$ is included in the supremum, we have 
\beq
Q_{M}(s) \geq 0, \qquad s>0. 
\eeq 
Using the operator-norm bound (\ref{eq:Knorm}) and the spectral symmetry of $K[U]$, we obtain
\beq
2M^3\log s^2
\leq\log\det D[s;U]
\leq2M^3\log\Big(s^2+\frac{9}{a^2}\Big).
\eeq
Applying these bounds to the numerator and denominator in Eq. (\ref{eq:ratio}), we obtain
\beq
0\leq Q_M(s)\leq 2\log\Big(1+\frac{9}{a^2s^2}\Big),\qquad s>0.
\eeq
Setting $s=\sqrt{m^2+t^2}$ in this inequality we obtain
\beq
0\leq Q_M\Big(\sqrt{m^2+t^2}\Big)\leq 2\log\Big(1+\frac{9}{a^2(m^2+t^2)}\Big).
\eeq
Since the upper bound is integrable in $[0, \infty)$, equation (\ref{eq:Qto0}) implies 
\beq
\lim_{M\to\infty}\int_0^\infty Q_M\Big(\sqrt{m^2+t^2}\Big)\,dt=0.
\eeq
Thus, using the integral representation (\ref{eq:integral_rep_1}), we obtain the desired result 
\beq
\lim_{N_{\Omega} \to\infty}
\frac{\displaystyle\min_U E_f^{(0)}[U]-E_f^{(0)}[I]}{N_{\Omega}}
=0.
\eeq
We emphasize that $N_\Omega=M^3$, where $M\geq3$ may be either even or odd. Thus, the result holds in the thermodynamic limit independently of lattice parity.
\subsection{Finite-Lattice Counterexamples}
The preceding result establishes the asymptotic optimality of the free-field configuration, but does not imply that it minimizes the fermionic ground-state energy at finite lattice size. Indeed, there exist nontrivial gauge configurations whose fermionic ground-state energy is strictly lower than that of the free configuration. We illustrate this distinction with an explicit counterexample on a periodic $3\times3\times3$ lattice. For simplicity, we take set $m=a =1$ and $q=-1.$ Let $\n = (n_1, n_2 ,n_3)\in\{-1,0,1\}^3$, with periodic boundary conditions. Consider the gauge potential
\beq
A_1(\mathbf n)=\frac{2\pi}{9}-2\pi\delta_{n_2,0}\delta_{n_3,0},
\qquad A_2(\mathbf n)=A_3(\mathbf n)=0.
\eeq
First, note that 
\beq
\sum_{\n \in \Omega_1} \A( \n) = 0, 
\eeq
and is therefore compatible with the exclusion of the zero-momentum photon mode. Moreover, since $A_1(\mathbf n)$ is independent of $n_1$ and $A_2=A_3=0$, the lattice Coulomb condition is also satisfied,
\beq
\nabla^-\cdot\mathbf A=0.
\eeq
The corresponding link variables are 
\beq
U_1(\mathbf n)=e^{2\pi i/9},\qquad U_2(\mathbf n)=U_3(\mathbf n)=1.
\eeq
Using the spectrum of the hopping matrix for constant link variables, we obtain
\beq
E_f^{(0)}[U]=-2\sum_{\boldsymbol{\ell} \in \{-1, 0, 1\}^3}\sqrt{1+\sum_{j=1}^{3}\sin^2\left(\frac{2\pi \ell_j}{3}-\theta_j\right)},
\eeq
where $\boldsymbol{\theta} = (2\pi/9, 0, 0).$
Evaluating the finite momentum sum, we obtain
\beq
E_f^{(0)}[U]-E_f^{(0)}[I]
\approx-0.0209093700066282<0.
\eeq
Thus, the free-field configuration does not minimize the fermionic ground-state energy on this finite lattice, despite the gauge potential satisfying the Coulomb condition and having no zero-momentum component.

Next, we consider a periodic $4\times4\times4$ lattice, again taking $m=a=1$ and $q=-1$. Let $\n=(n_1,n_2,n_3)\in \{0,1,2,3\}^3$, with periodic boundary conditions. Define
\beq
f(a , b)=\delta_{a,0}(\delta_{b,0}+\delta_{b,1}),
\eeq
and consider the gauge potential
\beq
\begin{aligned}
A_1(\mathbf n)&=\frac{\pi}{4}-2\pi f(n_2,n_3),\\
A_2(\mathbf n)&=\frac{\pi}{4}-2\pi f(n_3,n_1),\\
A_3(\mathbf n)&=\frac{\pi}{4}-2\pi f(n_1,n_2).
\end{aligned}
\eeq
By construction, 
\beq
\sum_{\n} \A(\n) =0, \qquad \nabla^- \cdot \A = 0. 
\eeq
The corresponding link variables are constant,
\beq
U_j(\mathbf n)=e^{i\pi/4},\qquad j=1,2,3.
\eeq
For this configuration, direct evaluation of the hopping matrix gives
\beq
K[U]^2=\frac{3}{2} I_{256}.
\eeq
Thus, the Cauchy–Schwarz bound (\ref{eq:cauchy_schwarz_bound}) is saturated, and this configuration attains the global minimum of the fermionic ground-state energy on this $4\times 4 \times 4$ lattice. Consequently, 
\beq
E_f^{(0)}[U]-E_f^{(0)}[I]
=48(1+\sqrt2+\sqrt3)-64\sqrt{10}
\approx -3.365080493562<0.
\eeq
\section{Fock-vacuum expectation values}
We now evaluate the Fock-vacuum expectation values of the Coulomb interaction, the fermion–gauge hopping term, and the fermion mass term. These quantities determine the reference energy that is subtracted from the interacting Hamiltonian, ensuring that the Fock vacuum has zero expectation value. We take the reference state to be the tensor product of the free fermionic and photon vacua,
\beq
\ket{0} = \ket{0_f} \otimes \ket{0_g}. 
\eeq
\subsection{The Coulomb expectation value}
We begin with the Coulomb contribution to the Fock-vacuum energy. Since the Coulomb Hamiltonian acts only on the fermionic sector, its expectation value is determined entirely by the free fermionic vacuum. Recall that in momentum space the Coulomb Hamiltonian takes the form
\beq
\hat H_c = \frac{q^2}{2 V} \sideset{}{'} \sum_{\q} \frac{\NO{\hat J^{\dagger}_0 (\q)}\NO{\hat J^{}_{0}(\q)}}{\tilde{\q}^2}. 
\label{eq:Coulomb_0}
\eeq
We introduce the Fock-vacuum charge-density correlation function
\beq
\mathcal{R}_0 (\q) := \bra{0_f}\NO{\hat J^{\dagger}_0 (\q)}\NO{\hat J^{}_{0}(\q)}\ket{0_f},
\label{eq:R0}
\eeq
where the Fourier transform of the charge-density operator is defined by
\beq
\hat J_0 (\q) = \Delta V \sum_{\r} \hat J_0 (\r) e^{- i \q \cdot  \r}.
\eeq

Using the index notation $i\equiv (\k, s)$ and $j \equiv (\p, s')$ we write 
\beq
\NO{\hat J_0 (\r)} = :\hat{\Psi}^{\dagger}(\r) \hat{\Psi}^{}(\r):=  \hat{b}^{\dagger} \mathcal{U}(\r)\, \hat{b} - \hat{d}^{\dagger} \mathcal{V}(\r)\,\hat{d} + \hat{b}^{\dagger}\mathcal{W}(\r)\,\hat{d}^{\dagger} + \hat{d}\, \mathcal{W}^{\dagger}(\r)\,\hat{b}, 
\eeq
where $\mathcal{U}(\r)$, $\mathcal{V}(\r)$ and $\mathcal W (\r)$ are $2 N_{\Omega}\times 2 N_{\Omega}$ matrices. With this notation, we have the vacuum correlator 
\beq
\bra{0_f} \NO{\hat J_0 (\r)} \NO{\hat J_0 (\r')}\ket{0_f}= \Tr[\mathcal W^{\dagger} (\r) \mathcal W^{}(\r')], 
\label{eq:Coulomb_W}
\eeq
where 
\beq
\mathcal{W}_{\k, s; \p, s'}(\r)=\frac{1}{V} \frac{m}{\sqrt{E_{\k} E_{\p}}}  u^{\dagger}(\bar{\k}, s) v(\bar{\p}, s')e^{- i (\k + \p) \cdot \r}.
\label{eq:W}
\eeq
Combining Eqs.~(\ref{eq:Coulomb_W}) and (\ref{eq:R0}) gives
\beq
\mathcal{R}_0 (\q) = (\Delta V)^2 \sum_{\r, \r'} \Tr[\mathcal W^{\dagger} (\r) \mathcal{W}(\r')]e^{i \q \cdot (\r- \r')}. 
\eeq
Using the matrix elements in  Eq.~(\ref{eq:W}), we obtain 
\beq
\mathcal{R}_0 (\q) = \sum_{\k, \p \in \Omega_N^{\star}} \frac{m^2}{E_{\k} E_{\p}} \sum_{s, s'} \abs{ u^{\dagger}(\bar \k, s) v(\bar \p, s')}^2 \delta^{\mathrm{BZ}}_{\k+ \p + \q, 0}. 
\label{eq:coulomb_2}
\eeq
Here the Brillouin-zone Kronecker delta is defined for momenta \(\k,\p\in(2\pi/L)\mathbb Z^3\) by
\beq
\delta^{\mathrm{BZ}}_{\k,\p}
:=
\frac{\Delta V}{V}
\sum_{\mathbf r\in\Omega_N}
e^{-i(\mathbf k-\mathbf p)\cdot\mathbf r}.
\eeq
Introducing the reciprocal-lattice translation vectors $\mathbf G\in\mathcal G=(2\pi/a)\mathbb Z^3$, we obtain
\beq
\delta^{\mathrm{BZ}}_{\k,\mathbf \p}
=
\begin{cases}
1,&\k-\p\in\mathcal G,\\
0,&\k-\mathbf p\notin\mathcal G.
\end{cases}
\eeq
When both arguments belong to the chosen Brillouin-zone momentum set \(\Omega_N^*\), this reduces to the ordinary Kronecker delta. In Eq. (\ref{eq:coulomb_2}), however, the sum \(\mathbf k+\mathbf p+\mathbf q\) can equal a nonzero reciprocal-lattice vector even though each momentum belongs to \(\Omega_N^*\). The Brillouin-zone delta therefore retains the Umklapp contributions, for which
\beq
\k+\p+\q=\mathbf G\ne\mathbf0,
\qquad \mathbf G\in\mathcal G.
\eeq
These terms would be omitted by imposing $\k+\p+\q=\mathbf0\) alone.
The periodic structure of the fermionic lattice momentum now simplifies the calculation. Since 
\beq
\bar k_j=\frac{1}{a} \sin(ak_j), 
\eeq
shifts by reciprocal-lattice vectors leave both the lattice momentum and the energy unchanged:
\[
\overline{\k+\mathbf G}=\overline{\k},
\qquad
E_{\k+\mathbf G}=E_{\k},
\qquad \mathbf G\in\mathcal G.
\]
Consequently, the constraint $\p=-(\k+\q)+\mathbf G$ implies
\beq
\overline{\p}=-\overline{\k+\q},
\qquad
E_{\p}=E_{\k+\q}.
\eeq
The reciprocal-lattice vector therefore drops out of the summand in Eq. (\ref{eq:coulomb_2}). For each $\k$ and $\q$, the Brillouin-zone delta selects a unique $\p\in\Omega_N^{\star}$, allowing us to eliminate the $\p$ sum directly. Normal and Umklapp contributions are thus included in a single expression. Thus, 
\beq
\mathcal R_{0}(\q) = \sum_{\k \in \Omega^{\star}_N}
\frac{m^2}{E_{\k} E_{\k + \q}}\sum_{s, s'}\abs{u^{\dagger}(\bar \k, s) v(- \overline{\k + \q}, s')}^2. 
\eeq
Using the spin-summation formula in Eq.~(\ref{ident_dirac_11}), we obtain 
\beq
\mathcal R_0 (\q) = \sum_{\k \in \Omega_N^{\star}} \frac{E_{\k} E_{\k + \q} - \bar{\k} \cdot  \overline{\k + \q} - m^2}{E_{\k} E_{\k + \q}}.
\label{eq:R0_1}
\eeq
Note that 
\beq
\mathcal R_0 (0) = 0,  
\eeq
as  required by the charge neutrality of the Fock vacuum. Using Eqs.~(\ref{eq:Coulomb_0}) and (\ref{eq:R0_1}), the Fock-vacuum expectation value of the Coulomb Hamiltonian becomes 
\beq
\bra{0_f}|\hat H_c \ket{0_f} = \frac{q^2}{2 V} \sideset{}{'}\sum_{\q \in \Omega_N^{\star}}\frac{1}{\tilde{q}^2}\sum_{\k \in \Omega_{N}^{\star}}\Big[1 - \frac{\bar \k \cdot \overline{\k + \q} +m^2}{E_{\k} E_{\k+\q}}\Big]. 
\eeq
Introducing the dimensionless lattice sums $C_0(N)$ and $C_1(N,\mu)$, we can express the above expression as 
\beq
\bra{0_f} \hat H_c \ket{0_f} = \frac{q^2 N_{\Omega}}{8 a} \Big( C_0 (N) - C_1 (N, \mu) \Big), 
\eeq
where $\mu = m a.$ The lattice sums are given by 
\beq
C_0 (N) = \frac{1}{N_{\Omega} } \sideset{}{'}\sum_{\q \in \Omega_N^{\star}} \frac{1}{\sum_{j=1}^3 \sin^2(q_j a/2)},
\label{eq:lattice_sum_C0_def}
\eeq
and 
\beq
C_1 (N, \mu) = \frac{1}{N_\Omega^2} \sideset{}{'}\sum_{\q \in \Omega^{\star}_N}\frac{1}{\sum_{j=1}^3 \sin^2(q_j a/2)} \sum_{\k \in \Omega_N^{\star}}\frac{\bar \k \cdot \overline{\k + \q} + m^2}{E_{\k} E_{\k + \q}}. 
\label{eq:lattice_sum_C1_def}
\eeq
\subsection{The fermion-gauge Hamiltonian expectation value}
We now turn to the Fock-vacuum expectation value of the fermion–gauge hopping Hamiltonian,
\beq
\bra{0}\Hfg\ket{0}.
\eeq
Using the lattice hopping Hamiltonian and the factorization of the Fock vacuum, we obtain
\beq
\bra{0}\Hfg \ket{0} = - i \frac{\Delta V}{2 a} \sum_{\r, i} \Big( \bra{0_f}\hat \Psi^{\dagger}(\r)\alpha^{i} \hat \Psi(\r + a \ei)\ket{0_f}\bra{0_g}\hat U_i^{\dagger}(\r)\ket{0_g} - \mathrm{c.c.}).
\eeq
For convenience, we introduce the photon link variable expectation value and the fermionic hopping correlator,
\beq
\mathcal{U}_i (\r) := \bra{0_g}\hat U_i (\r) \ket{0_g}, \qquad \mathcal{J}^{i} (\r)  := \bra{0_f}\hat \Psi^{\dagger}(\r)\alpha^{i} \hat \Psi(\r + a \ei)\ket{0_f}. 
\label{eq:evs_fock_vac}
\eeq
Since the Fock vacuum is translationally invariant, both expectation values in Eq.~(\ref{eq:evs_fock_vac}) are independent of $\mathbf r$. Thus,
\beq
\mathcal{U}_i (\r)  = \mathcal{U}_i, \qquad \mathcal{J}^i (\r)  = \mathcal{J}^i,  
\eeq
and 
\beq
\bra{0}\Hfg\ket {0}= - i \frac{V}{2 a} \sum_i  \Big(\mathcal{J}^i \mathcal{U}^{*}_i - \mathrm{c.c.}\Big). 
\label{eq:ev_fg_1} 
\eeq
Now, since the photon vacuum is Gaussian and $\bra{0_g}\hat A_i (\r) \ket{0_g}=0$, we can use the identity 
\beq
\bra{0_g} e^{i \lambda \hat A_i}\ket{0_g}= \exp(- \frac{\lambda^2}{2} \bra{0_g}\hat A_i^2 \ket{0_g}). 
\eeq
Consequently, 
\beq
\mathcal{U}_i = \exp( - \frac{q^2 a^2}{2} \bra{0_g}\hat{A}^2_i (\r)\ket{0_g}).
\eeq
Using the mode expansion (\ref{eq:A_field_expansion}) and the completeness relation Eq.~(\ref{eq:completness_pol}),  we obtain 
\beq
 \bra{0_g}\hat{A}^2_i (\r)\ket{0_g}= \frac{1}{2 V} \sideset{}{'}\sum_{\k, \sigma} \frac{\abs{\epsilon_i(\k, \sigma)}^2}{\omega_k} = \frac{1}{2V}\sideset{}{'} \sum_{\k} \frac{1}{\omega_k} \big(1 - \frac{\tilde{k}_i^2}{\tilde k^2}\big). 
\eeq
Since the momentum lattice is invariant under permutations of the coordinate axes, and $\omega_{\mathbf k}=|\tilde{\k}|$, we have
\beq
\sideset{}{'} \sum_\k \frac{\tilde{k}_i^2}{\tilde k^3} = \frac{1}{3} \sideset{}{'}\sum_\k  \frac{1}{ \tilde k}. 
\eeq
Consequently 
\beq
 \bra{0_g} |\hat{A}^2_i (\r)\ket{0_g} = \frac{1}{3 V} \sideset{}{'} \sum_{\k} \frac{1}{\tilde k}, 
\eeq
and 
\beq
\mathcal{U}_i =  \exp( - \frac{q^2 a^2}{6 V}  \sideset{}{'} \sum_{\k} \frac{1}{\tilde k}).
\label{eq:C2_def}
\eeq
Note that 
\beq
\mathcal{U}_1 = \mathcal{U}_2 = \mathcal{U}_3 = \mathcal{U}. 
\eeq
Moreover, since $\mathcal{U}$ is real, the expression (\ref{eq:ev_fg_1}) becomes 
\beq
\bra{0}\Hfg\ket{0}= - i \frac{V}{2 a}\mathcal{U} \sum_i  \Big(\mathcal{J}^i  - \mathrm{c.c.}\Big). 
\eeq

Now, using the Dirac-field mode expansion in Eq.~(\ref{eq:dirac_field_expansion}), we obtain 
\beq
\mathcal{J}^i = \frac{m}{V} \sum_{\k, s} \frac{1}{E_k} v^{\dagger}(\bar \k, s) \alpha^{i} v(\bar \k, s) e^{- i k^i a}. 
\eeq
The spinor identity (\ref{ident_dirac_2}) yields 
\beq
\mathcal{J}^i = \frac{2}{V} \sum_{\k} \frac{1}{E_k} \bar k^i e^{- i k^i a}  = -\frac{2 i a}{V} \sum_{\k} \frac{(\bar k^{i})^2}{E_k}. 
\eeq
Exploiting the cubic symmetry of the lattice once more, we find
\beq
\mathcal{J}^i = - \frac{2 i a}{3 V} \sum_{\k} \frac{\bar k^2}{E_k}. 
\eeq
The Fock-vacuum expectation value is therefore
\beq
\bra{0}\Hfg \ket{0} =- 2\, \mathcal{U} \sum_{\p \in \Omega_N^{\star}}  \frac{\bar p^2}{E_{\p}} <0. 
\eeq
Note that the Fock-vacuum expectation value of the gauge-covariant hopping Hamiltonian is strictly negative.
\subsection{The mass term Hamiltonian}
\beq
\hat H_m = m \Delta V \sum_{\r} \hat \Psi^{\dagger}(\r) \beta \hat \Psi(\r). 
\eeq
Using the Dirac-field mode expansion in Eq.~(\ref{eq:dirac_field_expansion}) and the spinor normalization (\ref{eq:v_norm_2}), we find 
\beq
\bra{0_f} \hat H_m \ket{0_f} = - 2 m^2 \sum_{\k \in \Omega^{\star}_{N}}\frac{1}{E_{\k}}. 
\eeq
We can then combine the mass and fermion–gauge contributions:
\beq
\bra{0}\hat H_m+\Hfg \ket{0}=-2\sum_{\k\in\Omega_N^*}\frac{m^2+\mathcal U \; \bar{\k}^{\,2}}{E_{\k}}.
\eeq
As a consistency check, consider the limit $\mathcal U\to1$. Using $E_{\k}^2=m^2+\bar{\k}^{2}$, the above equation reduces to 
\beq
\bra{0}\hat H_m+ \Hfg \ket{0}\rightarrow -2 \sum_{\k \in \Omega_N^{\star}} E_{\k}, 
\eeq
which is precisely the free-fermion Fock-vacuum energy.
\section{Lattice Sums}
\subsection{Coulomb lattice sums $C_0$ and $C_1$}
We begin with the lattice sum \(C_0(N)\), introduced in Eq. (\ref{eq:lattice_sum_C0_def}), which can be written explicitly as
\beq
C_0(N)
=
\frac{1}{(2N+1)^3}
\sum_{\substack{\mathbf n\in\{-N,\ldots,N\}^3\\ \mathbf n\ne0}}
\frac{1}{
\sum_{i=1}^3
\sin^2\!\left(\frac{\pi n_i}{2N+1}\right)
}.
\label{eq:lattice_sum_C0_1}
\eeq
We define 
\beq
x_j=\frac{n_j}{2N+1},
\qquad \Delta x=\frac{1}{2N+1}.
\eeq
In the thermodynamic limit $N\to\infty$, the sum in Eq.~(\ref{eq:lattice_sum_C0_1}) converges to the integral
\beq
C_0(\infty)=\int_{[-1/2,1/2]^3}\frac{d^3\mathbf x}{\sum_{j=1}^3\sin^2(\pi x_j)}.
\eeq
Upon the substitution $t_j=2\pi x_j$, this integral becomes 
\beq
C_0(\infty)=\frac{2}{\pi^3}\int_{[0,\pi]^3}\frac{d^3 \mathbf t}{3-\cos t_1-\cos t_2-\cos t_3} = 2 W_s, 
\label{eq:C0_Watson}
\eeq
where $W_s$ is the simple-cubic Watson integral \cite{zucker201170+, doyle1984random}. The Watson integral admits the closed-form evaluation
\beq
W_s=\frac{\sqrt3-1}{96\pi^3}\Gamma^2\left(\frac1{24}\right)\Gamma^2\left(\frac{11}{24}\right).
\label{eq:Watson_integral}
\eeq
Consequently,
\beq
C_0(\infty)=2W_s=\frac{\sqrt3-1}{48\pi^3}\Gamma^2\left(\frac1{24}\right)\Gamma^2\left(\frac{11}{24}\right)\approx1.01092.
\label{eq:C0_num}
\eeq

We now turn to the lattice sum $C_1(N,\mu).$ In the thermodynamic limit $N\to\infty$, Eq.~(\ref{eq:lattice_sum_C1_def}) becomes
\beq
C_1(\infty,\mu)=\int_{[-1/2,1/2]^3}\!\!d^3 \mathbf x\int_{[-1/2,1/2]^3}\!\!d^3\mathbf y\,
\frac{\mu^2+\sum_{j=1}^3\sin\big(2\pi x_j\big)\sin\big(2\pi(x_j+y_j)\big)}
{\Big[\sum_{j=1}^3\sin^2(\pi y_j)\Big]D(\mathbf x, \mu)D(\mathbf{x+y}, \mu)},
\label{eq:C1_1}
\eeq
where 
\beq
D(\mathbf x, \mu)=\sqrt{\mu^2+\sum_{j=1}^3\sin^2(2\pi x_j)}.
\eeq
The integral retains the full dependence on the dimensionless fermion mass $\mu=ma$. We introduce the normalized four-component function
\beq
\mathbf f (\mathbf x, \mu)=\frac{\big[\mu,\sin(2\pi x_1),\sin(2\pi x_2),\sin(2\pi x_3)\big]}{D(\mathbf x, \mu)}, \qquad \mathbf f (\mathbf x, \mu) \cdot \mathbf  f(\mathbf x, \mu)=1. 
\eeq
Using this notation, Eq.~(\ref{eq:C1_1}) takes the form
\beq
C_1(\infty,\mu)=
\int_{[-1/2,1/2]^3}
\frac{d^3\mathbf y}{\sum_{j=1}^3\sin^2(\pi y_j)}
\int_{[-1/2,1/2]^3}
d^3\mathbf x\;
\mathbf f (\mathbf x, \mu)\cdot\mathbf f(\mathbf{x+y}, \mu).
\label{eq:C1_2}
\eeq
We expand the normalized vector field $\mathbf f(\mathbf x,\mu)$ in Fourier modes,
\beq
\mathbf f(\mathbf x,\mu)
=\sum_{\mathbf n\in\mathbb Z^3}
\mathbf F(\mathbf n,\mu)e^{2\pi i\mathbf n\cdot\mathbf x},
\eeq
where 
\beq
\mathbf F(\mathbf n,\mu)
=\int_{[-1/2,1/2]^3}
\mathbf f(\mathbf x,\mu)e^{-2\pi i\mathbf n\cdot\mathbf x}\; d^3\mathbf x. 
\label{eq:F_1_def}
\eeq
Because $\mathbf f$ is real,
\beq
\mathbf F(-\mathbf n,\mu)=\mathbf F(\mathbf n,\mu)^*.
\label{eq:reality}
\eeq
Using Fourier orthogonality and the reality condition (\ref{eq:reality}), the inner integral in (\ref{eq:C1_2}) becomes
\beq
\int_{[-1/2,1/2]^3}\mathbf f(\mathbf x,\mu)\cdot\mathbf f(\mathbf{x+y},\mu)\; d^3\mathbf x 
=
\sum_{\mathbf n\in\mathbb Z^3}
|\mathbf F(\mathbf n,\mu)|^2
e^{2\pi i\mathbf n\cdot\mathbf y}.
\eeq
Substituting this relation into $(\ref{eq:C1_2})$, we obtain 
\beq
C_1 (\infty, \mu) = \sum_{\n \in \mathbb{Z}^3} G(\n) \abs{F(\n, \mu)}^2, 
\label{eq:C_1_useful}
\eeq
where 
\beq
G(\n)=
\int_{[-1/2,1/2]^3}
\frac{e^{2\pi i\n\cdot\mathbf y}}
{\sum_{j=1}^3\sin^2(\pi y_j)} d^3\mathbf y. 
\label{eq:G(n)_1}
\eeq
Using the substitution $t_j=2\pi y_j$, 
\beq
G(\mathbf n)=\frac{2}{(2\pi)^3}
\int_{[-\pi,\pi]^3}
\frac{e^{i\mathbf n\cdot\mathbf t}}
{3-\sum_{j=1}^3\cos t_j}\,d^3\mathbf t.
\eeq
Now, using the integral representation 
\beq
\frac{1}{A} = \int_{0}^{\infty} e^{-s A}\; d s, \qquad A> 0, 
\eeq
we can rewrite the above integral as 
\beq
G(\n) = \frac{2}{(2\pi)^3} \int_{0}^{\infty} e^{- 3 s} \prod_{j=1}^{3}\Big(\int_{-\pi}^{\pi} e^{s \cos(t_j)} e^{i n_j t_j}\;dt_j  \;\Big) ds. 
\eeq
Next, using the integral representation of the modified Bessel function of the first kind 
\beq 
I_{n}(s) = \frac{1}{2\pi} \int_{-\pi}^{\pi} e^{s \cos (\theta)}  e^{i n \theta} \; d\theta, 
\eeq	
we obtain 
\beq
G(\n) = 2 \int_{0}^{\infty} e^{-3 s} I_{n_1}(s) I_{n_2}(s) I_{n_3}(s)\; ds. 
\label{eq:G(n)_2}
\eeq
Note from Eq.~(\ref{eq:G(n)_1}) that 
\beq
G(\mathbf 0) = 2 \int_{0}^{\infty} e^{-3 s} I^3_{0}(s)\; ds= 2 W_s, 
\eeq
where $W_s$ is the Watson integral Eq.~(\ref{eq:C0_Watson}). Setting $\mu=0$ and $t_i = 2 \pi x_i$ in Eq.~(\ref{eq:F_1_def}), we obtain 
\beq
F_j(\n, 0) = \int_{[-\pi, \pi]^3} \frac{\sin(t_j)e^{- i \n \cdot \mathbf t }}{\sqrt{\sum_{k=1}^3 \sin^2 (t_k)}}  \; \frac{d^3 \mathbf{t}}{(2\pi)^3}. 
\label{eq:Fj}
\eeq
Invoking the identity 
\beq
\frac{1}{\sqrt{A}} = \frac{1}{\sqrt{\pi}}\int_{0}^{\infty} \frac{e^{-s A}}{\sqrt{s}} d s, \qquad A>0,  
\eeq
we rewrite Eq.~(\ref{eq:Fj}) as 
\beq
F_j(\n, 0) = \frac{1}{\sqrt{\pi}} \int_{0}^{\infty} \frac{ds}{\sqrt{s}} \Big[\prod_{k \neq j} \int_{-\pi}^{\pi} \frac{d t_k}{2\pi}\exp(- s \sin^2 (t_k) - i n_k t_k)\Big]\times \Big[\int_{-\pi}^{\pi} \frac{d t_j}{2\pi} \sin(t_j) \exp(-s \sin^2 t_j - i n_j t_j)\Big]. 
\eeq
We use 
\beq
\int_{- \pi}^{\pi}  \frac{dt}{2\pi} e^{-s \sin^2 t - i n t} = \begin{cases} 
e^{-s/2} I_{n/2} (s/2), \qquad &n \; \mathrm{even} \\
0 &n \; \mathrm{odd} \\
\end{cases}
\eeq
and 
\beq
\sin( t_j) e^{- i n_j t_j} = \frac{e^{- i (n_j - 1) t_j} - e^{-i (n_j +1) t_j}}{2 i}. 
\eeq
Thus, for odd $n_j$ and even $n_k$, $k \neq j$, we have 
\beq
F_j (\n, 0) = \frac{1}{2 i \sqrt{\pi}}\int_{0}^{\infty} \frac{e^{- 3 s/2}}{\sqrt{s}} \times \Big[I_{(n_j - 1)/2} (\frac{s}{2}) - I_{(n_j + 1)/2}(\frac{s}{2}) \Big]\times \prod_{k \neq j} I_{n_k/2}(s/2)\; ds. 
\eeq
For all other combinations of parities, $F_j (\n, 0)$ vanishes. By cubic symmetry, the three spatial components of the Fourier coefficients give equal contributions to Eq.~(\ref{eq:C_1_useful}). Using the parity conditions established above, we therefore obtain
\beq
C_1 (\infty, 0) = 3 \sum_{\substack{\n \in \mathbb{Z}^3\\
n_1 \; \mathrm{odd}\\
n_2,\; n_3 \; \mathrm{even}}} G(\n) |F_1 (\n, 0)|^2. 
\label{eq:Fourier_sum_1}
\eeq
Exploiting the symmetry of the Fourier coefficients under sign changes of their indices, we can restrict the sum to nonnegative integers. Equation (\ref{eq:Fourier_sum_1}) then becomes
\beq
C_1 (\infty, 0) =  6 \sum_{r, p, q=0}^{\infty} w_p w_q \; G(2r +1, 2 p, 2 q) \abs{F_1 (2 r+1, 2 p, 2 q; 0)}^2, 
\label{eq:Fourier_sum_2}
\eeq
where 
\beq
w_p = 2 - \delta_{p 0}. 
\eeq
For numerical evaluation, we truncate the Fourier series in Eq. (\ref{eq:Fourier_sum_2}) to a finite region defined by
\beq
\norm{\n}_{\infty} = \max(|n_1|, |n_2|, |n_3|) \leq R. 
\eeq
To facilitate the numerical evaluation of $G(\n)$, we map the semi-infinite integration interval in Eq. (S197) to the finite interval \(0\leq u<1\) using the substitution
\beq
s=\frac{u^2}{(1-u)^2},\qquad 0\leq u<1,
\eeq
and the exponentially scaled modified Bessel functions
\beq
\widetilde I_n(s):=e^{-s}I_n(s).
\eeq
With these definitions, Eq. (\ref{eq:G(n)_2}) becomes
\beq
G(\mathbf n)=4\int_0^1
\frac{u}{(1-u)^3}
\prod_{j=1}^3
\widetilde I_{n_j}\left(\frac{u^2}{(1-u)^2}\right)\,du.
\eeq
The advantage of this representation is that the transformed integrand has a finite limit as \(u\to1^{-}\). Indeed,
\beq
\lim_{u\to1^-}
\frac{4u}{(1-u)^3}
\prod_{j=1}^3\widetilde I_{n_j}
\left(\frac{u^2}{(1-u)^2}\right)
=\frac{4}{(2\pi)^{3/2}}.
\eeq
The numerical values of the truncated sum \(C_1^{(R)}\) for increasing Fourier cutoff \(R\) are reported in Table~\ref{tab:C1}. The observed convergence suggests
\begin{table}[!t]
\centering
\renewcommand{\arraystretch}{1.5}
\setlength{\tabcolsep}{20pt}
\begin{tabular}{cc}
\hline\hline
$R$ & $C_1^{(R)}$ \\
\hline
10 & 0.333126780757 \\
15 & 0.333130008952 \\
20 & 0.333130444853 \\
25 & 0.333130603179 \\
\hline\hline
\end{tabular}
\caption{Convergence of the truncated lattice sum
$C_1^{(R)}$ with increasing Fourier cutoff $R$.}
\label{tab:C1}
\end{table}
\beq
C_1(\infty, 0) = 0.3331. 
\eeq
Combining this estimate with Eq.~(\ref{eq:C0_num}), we obtain
\beq
C_0(\infty)-C_1(\infty,0)\approx0.6778.
\eeq
\subsection{Lattice sums $C_2$ and $C_3$}
We first consider the dimensionless photon lattice sum appearing in the Fock-vacuum expectation value of the link variable, Eq.~(\ref{eq:C2_def}),
\beq
C_2 (N):= \frac{2 a^2}{V}\sideset{}{'}\sum_{\k\in \Omega_N^{\star}}  \frac{1}{\tilde k}.
\eeq
In terms of this sum, Eq. (\ref{eq:C2_def}) becomes \(\mathcal U=\exp(-q^2 C_2(N)/12)\). To evaluate its thermodynamic limit, we introduce the generalized Watson integral
\beq
W(r) := \frac{1}{\pi^3} \int_{[0, \pi]^3} \frac{ d^3 \mathbf t}{\big(3 - \cos (t_1) - \cos (t_2) - \cos (t_3)\big)^r} 
\eeq
This integral is finite for $r < 3/2,$ and the simple-cubic Watson integral introduced in Eq.~(\ref{eq:C0_Watson}) corresponds to $W(1) = W_s.$ Taking the thermodynamic limit and expressing the resulting integral in terms of $W(r)$, we obtain
\beq
C_2 (\infty) = \sqrt{2} W(\frac{1}{2}) \approx 0.9107.
\eeq

We finally introduce the dimensionless fermionic hopping
lattice sum
\begin{equation}
C_3(N,\mu):=\frac{2a}{N_\Omega}
\sum_{\mathbf k\in\Omega_N^\star}
\frac{\bar{\mathbf k}^{\,2}}{E_{\mathbf k}},
\qquad \mu=ma.
\end{equation}
In terms of this sum, the Fock-vacuum expectation value
of the hopping Hamiltonian becomes
\begin{equation}
\langle0|\hat H_{f-g}|0\rangle
=-\frac{N_\Omega}{a}\,\mathcal{U} C_3(N,\mu).
\end{equation}
In the thermodynamic limit, we obtain
\begin{equation}
C_3(\infty,\mu)=
2\int_{[-1/2,1/2]^3}
\frac{\sum_{j=1}^3\sin^2(2\pi x_j)}
{\sqrt{\mu^2+\sum_{j=1}^3\sin^2(2\pi x_j)}}
\,d^3\mathbf x.
\end{equation}
In the limit $\mu\to0$, the integral can be expressed in terms of the generalized Watson integral,
\begin{equation}
C_3(\infty,0)
=\sqrt{2}\,W(-1/2)\approx 2.3876.
\end{equation}
For small $\mu$, we find
\begin{equation}
C_3(\infty,\mu)
=C_3(\infty,0)-C_2(\infty)\mu^2+o(\mu^2).
\end{equation}

\end{document}